\documentclass[11pt,a4paper]{article}
\pdfoutput=1
\usepackage{jheppub}

\usepackage{amsfonts}
\usepackage{mathrsfs}
\usepackage{graphicx}
\usepackage{amssymb}
\usepackage{color}
\usepackage{amsmath}

\usepackage{amsfonts}
\usepackage{epstopdf}
\usepackage{slashed}
\usepackage{extarrows}

\newcommand{\td}{\text{d}}

\newcommand{\el}{\ell_{\text{AdS}}}

\usepackage{multirow, graphicx,amssymb,url,mathrsfs,amsmath}
\usepackage{wrapfig,boxedminipage,setspace,subfigure,epsfig}
\usepackage{amsxtra,amstext,latexsym,dsfont,amsfonts}
\usepackage{color,eucal}
\usepackage[dvipsnames]{xcolor}
\usepackage{float}

\title{Complexity of Holographic Superconductors}

\author[a]{Run-Qiu Yang,}
\author[b]{Hyun-Sik Jeong,}
\author[c]{Chao Niu,}
\author[b]{Keun-Young Kim,}

\emailAdd{aqiu@kias.re.kr}
\emailAdd{hyunsik@gist.ac.kr}
\emailAdd{chaoniu09@gmail.com}
\emailAdd{fortoe@gist.ac.kr}

\affiliation[a]{Quantum Universe Center, Korea Institute for Advanced Study, Seoul 130-722, Korea}
\affiliation[b]{School of Physics and Chemistry, Gwangju Institute of Science and Technology, Gwangju 61005, Korea}
\affiliation[c]{Department of Physics and Siyuan Laboratory, Jinan University, Guangzhou 510632, China}

\abstract{
We study the complexity of holographic superconductors (Einstein-Maxwell-complex scalar actions in $d+1$ dimension) by the ``complexity = volume'' (CV) conjecture. First, it seems that there is a universal property: the superconducting phase always has a smaller complexity than the unstable normal phase below the critical temperature, which is similar to a free energy. We investigate the temperature dependence of the complexity. In the low temperature limit, the complexity (of formation) scales as $T^\alpha$, where $\alpha$ is a function of the complex scalar mass $m^2$, the $U(1)$ charge $q$, and dimension $d$. In particular, for $m^2=0$, we find $\alpha=d-1$, independent of $q$, which can be explained by the near horizon geometry of the low temperature holographic superconductor. Next, we develop a general numerical method to compute the {\it time-dependent} complexity by the CV conjecture. By this method, we compute the time-dependent complexity of holographic superconductors. In both normal and superconducting phase, the complexity increases as time goes on and the growth rate saturates to a temperature dependent constant.   The higher the temperature is, the bigger the growth rate is. However, the growth rates do not violate the Lloyd's bound in all cases and saturate the Lloyd's bound in the high temperature limit at a late time.
}

\begin{document}

\maketitle

\section{Introduction}

The quantum information theory has been playing an important role in investigating the quantum gravity and quantum field theory.
The AdS/CFT correspondence or gauge/gravity duality set this idea in a holographic framework: some  geometric quantities in the bulk spacetime is related to the entanglement properties of the boundary field theory.
One of the most studied ideas is the entanglement entropy and the Ryu-Takayanagi formula~\cite{Ryu:2006bv, Ryu:2006ef}, from which the spacetime geometry may emerge (e.g., see~\cite{Nishioka:2009un,VanRaamsdonk:2010pw,Nozaki:2012zj,Lin:2014hva,Hayden:2016cfa}).  However, it does not cover the spacetime region inside the black hole horizon.  As a candidate to explore inside the black hole horizon, another important concept, the ``complexity'', has been introduced from quantum computation theory~\cite{Harlow:2013tf,Stanford:2014jda,Susskind:2014rva,Brown:2015bva,Brown:2015lvg,Czech:2017ryf}.

In a {\it discrete} quantum circuit system, the complexity counts how many fundamental gates are required to obtain a particular state of interest from a reference state \cite{Watrous:2008aa, Aaronson:2016vto, Gharibian:2014aa, Osborne:2011aa}.  Intuitively, it measures how difficult it is to obtain a particular target state from a certain reference state. Recently, there have been many attempts to generalize the concept of complexity of discrete quantum circuit to continuous systems such as ``complexity geometry''~\cite{Susskind:2014jwa,Brown:2016wib,Brown:2017jil} based on ~\cite{Nielsen:2006aa,Nielsen:aa,Dowling:aa}, Fubini-study metric~\cite{Chapman:2017rqy}, and path-integral optimization~\cite{Caputa:2017urj,Caputa:2017yrh,Bhattacharyya:2018wym,Takayanagi:2018pml}. See also \cite{Hashimoto:2017fga,Hashimoto:2018bmb,Flory:2018akz,Flory:2019kah,Belin:2018fxe,Belin:2018bpg}~\footnote{In particular, the complexity geometry is
the most studied. See for exampe~\cite{Jefferson:2017sdb,Yang:2017nfn,Reynolds:2017jfs,Kim:2017qrq,Khan:2018rzm,Hackl:2018ptj,Yang:2018nda,Yang:2018tpo,Alves:2018qfv,Magan:2018nmu,Caputa:2018kdj,Camargo:2018eof,Guo:2018kzl,Bhattacharyya:2018bbv, Jiang:2018gft,Camargo:2018eof,Chapman:2018hou,Ali:2018fcz,Chapman2018}.}.
From the holographic perspective, there are two widely studied proposals to compute the complexity in holography\footnote{There are also other holographic proposals for complexity, see Refs.~\cite{Alishahiha:2015rta,Ben-Ami:2016qex,Couch:2016exn,Fan:2018wnv} for examples.}: the CV (complexity=volume) conjecture~\cite{Susskind:2014rva,Stanford:2014jda,Alishahiha:2015rta} and the CA (complexity= action) conjecture~\cite{Brown:2015bva,Brown:2015lvg}.

In the CV conjecture, the complexity of a state {$|\psi(t_L,t_R)\rangle$} corresponds to the maximal volume of the codimension-one surface connecting the codimension-two time slices (denoted by $t_L$ and $t_R$) at two AdS boundaries:
\begin{equation}\label{CV}
  \mathcal{C}_V=\max_{\partial \Sigma=t_L\cup t_R}\left[\frac{V(\Sigma)}{G_N \ell}\right] \,,
\end{equation}
where $\Sigma$ is all possible codimension-one surface connecting $t_L$ and $t_R$, $G_N$ is the Newton's constant, and $\ell$ is some length scale such as  the horizon, AdS radius or something related with the bulk geometry.
In the CA conjecture, the complexity of a state {$|\psi(t_L,t_R)\rangle$} corresponds to the action in the Wheeler-DeWitt (WDW) patch
\begin{equation}\label{CA}
  \mathcal{C}_A=\frac{I_{\text{WDW}}}{\pi\hbar}.
\end{equation}
where the WDW patch is the collection of all space-like surface connecting $t_L$ and $t_R$ bounded by the null sheets coming from $t_L$ and $t_R$. 
%
%
There have been many works on holographic complexity.
For example, see~\cite{Susskind:2014rva, Stanford:2014jda, Roberts:2014isa,Brown:2015bva,Brown:2015lvg,Cai:2016xho,Lehner:2016vdi,Chapman:2016hwi,Carmi:2016wjl,Reynolds:2016rvl,Kim:2017lrw,Carmi:2017jqz,Kim:2017qrq,Swingle:2017zcd,Alishahiha:2018tep,Alishahiha:2018lfv,Fan:2018xwf,Moosa:2017yvt}.

In this paper, we analyze the complexity of the holographic superconductors for the first time.  For our method, we only focus on the CV conjecture, leaving the CA conjecture and field theoretic methods as future research directions. {A main difficulty of the CA conjecture is due to the region of the singularity involved in the on-shell action computation. In the cases where analytical solutions are available, we may deal with the singularity easily. Otherwise, with only numerical solutions, it is subtle and difficult to study the geometry and matter fields around the singularity. Such a difficulty does not appear in the CV conjecture because the extremal hyper-surface will not touch the singularity. Thus, in this paper, we only investigate the CV conjecture.} Our main subjects are i) the complexity of formation and ii) the time-dependent complexity.

First, the complexity of formation was defined in \cite{Chapman:2016hwi} by the complexity of a thermal state such as the AdS$_{d+1}$-Schwarzschild or AdS$_{d+1}$-RN blackhole geometry from the pure AdS$_{d+1}$ spacetime at $t=0$. For the AdS$_{d+1}$-Schwarzschild case, the complexity of formation scales as $T^{d-1}$. For the AdS$_{d+1}$-RN case, it scales as $T^{d-1}$ in the high temperature limit and $\ln T$ in the low temperature limit.  In this paper, as another thermal state,  we consider holographic superconductors.  We find that the change of the complexity of formation at the critical temperature is not smooth. In superconducting phase, in the \textit{low} temperature limit, the complexity of formation scales as $T^\alpha$, where $\alpha$  is a function of the complex scalar mass $m^2$, the $U(1)$ charge $q$, and dimension $d$. It has something to do with the near horizon geometry of the low temperature holographic superconductor.
In particular, for $m^2=0$ we find $\alpha=d-1$, independent of $q$. It is due to the fact that, in the zero temperature limit, the near horizon geometry of the holographic superconductor with $m^2=0$ is just the AdS$_{d+1}$-Schwarzschild.

Next, we compute the full time evolution of holographic complexity.
The holographic time-dependent complexity for the AdS$_{d+1}$-Schwarzschild or AdS$_{d+1}$-RN blackhole geometry has been studied in \cite{Carmi:2017jqz}. These works focus on the late time behavior and rely on the analytic solutions. To compute the full time evolution for the holographic superconductor, we first develop a general numerical method to compute time-dependent complexity. By using this method we compute the complexity of the holographic superconductor in both normal and superconducting phase. The complexity always increases as time goes on and the growth rate saturates to a temperature dependent constant.  The higher the temperature is, the bigger the growth rate is. {However, the growth rates do not violate the Lloyd's bound (intuitively meaning the fastest computing time~\cite{Lloyd_2000,Markov_2014,Hartman2016hs})}
n all cases and saturates the Lloyd's bound in the high temperature limit at late time. This result is a nontrivial support for the conjecture that the Schwarzshield black is the fastest quantum computer of the same energy.   Let us recall that the absolute value of the complexity growth rate in the CA conjecture can be arbitrarily large and does not satisfy the Lloyd's bound~\cite{Carmi:2017jqz,Kim:2017qrq}.\footnote{In the late time limit, the complexity growth rate satisfies the Lloyd's bound if matter fields satisfy some requirements, e.g., see Ref.~\cite{Yang:2016awy}.} {Thus, our results indicate that, from the perspective of the Lloyd's bound, the CV conjecture is more suitable than the CA conjecture as a definition of the holographic complexity.}

The paper is organized as follows: In section 2, we briefly review our holographic superconductor model (Einstein-Maxwell-complex scalar action). In section 3, using the CV conjecture, we study the time-independent complexity of formation of the holographic superconductor. In particular, we investigate the low temperature behavior of the  complexity of formation in both normal and superconducting phase. In section 4, we develop a general method to compute {\it time dependent }holographic CV conjecture. By this method, we analyze the full time evolution of the complexity of the holographic superconductor, from normal to superconducting phase. We investigate the growth rate of the complexity and compare it with the Lloyd's bound. We conclude in section 5.


\section{Holographic superconductor model: a quick review}

Let us first introduce a holographic superconductor model that we consider in this paper. It is the first holographic superconductor model proposed by Hartnoll, Herzog, and Horowitz~\cite{Hartnoll:2008vx,Hartnoll:2008kx}. Here, we describe only essential features of the model necessary for our study and refer to~\cite{Hartnoll:2008vx,Hartnoll:2008kx} for more details. The action reads
\begin{eqnarray}\label{SetupModel}
	I_{\text{bulk}}=\frac1{16\pi}\int
	\td^{d+1}x\sqrt{-g} \left[ R+\frac{d (d-1)}{\ell_{\text{AdS}}^2}-\frac{1}{4}F^2 -|D\Phi|^2-m^2|\Phi|^2 \right]\,, \label{eq:action}
\end{eqnarray}
where $D=\nabla-iqA$. $F=\td A$ is a field strength of a  bulk $U(1)$ gauge field $A$ and encodes a finite chemical potential or density in field theory.  $\Phi$ is a complex scalar field with mass $m$ and has something to do with the order parameter of the superconducting phase transition. From here, we set  $\ell_{\text{AdS}}=1$. The equations of motion obtained from this action are
\begin{align}
&\nabla^\mu  F_{\mu\nu} = i q ( \Phi^* D_\nu \Phi - \Phi D_\nu \Phi^*   ) \,, \label{eom1} \\
&\left(D^2 - m^2 \right) \Phi =0 \,, \label{eom2} \\
& G_{\mu\nu} - \frac{d(d-1)}{2}g_{\mu\nu}= \frac{1}{2} T_{\mu\nu}  \,, \label{eom3}
\end{align}
where $T_{\mu\nu}$ is the energy-momentum tensor given by
\begin{eqnarray}
T_{\mu\nu}=F_{\mu\sigma}{F_\nu}^\sigma + \left(  D_\mu \Phi D_\nu \Phi^* + D_\nu \Phi D_\mu \Phi^* \right)+g_{\mu\nu} \left(  -\frac{1}{4}F^2   -  |D\Phi|^2 - m^2  |\Phi| ^2  \right).
\end{eqnarray}

If we consider the following ansatz for metric and matter fields
\begin{equation}\label{anstz1}
\begin{split}
\td s^2=&{1\over
	z^2}\left[-f(z)e^{-\chi(z)}\td t^2+f^{-1}(z)\td z^2  +     \sum_{i=1}^{d-1}\td x^2_i      \right], \\
A = &A_t(z) \td t\,, \qquad  \Phi= \phi(z)\,,
\end{split}
\end{equation}
with four functions; $f(z), \chi(z), A_t(z)$ and $\phi(z)$,
the equations of motion \eqref{eom1}-\eqref{eom3} boil down to
\begin{equation}\label{eqsfchiAphi}
\begin{split}
0 & = \chi'  - \frac{2}{d-1}\left( \frac{z \, q^{2} A_{t}^2 \phi^2 e^{\chi}}{f^2} + z \, \phi'^2 \right)  \,, \\
0 & = f' - \left( \frac{d}{z} + \frac{z \, \phi'^2}{d-1}\right)f   - \frac{1}{d-1} \left(  \frac{z^3 A_{t}'^2 e^{\chi}}{2}   +  \frac{m^2 \phi^2}{z}    +   \frac{z \, q^2 A_{t}^2 \phi^2 e^{\chi}}{f} \right)  + \frac{d}{z} \,, \\
0 & = A_{t}'' - \left( \frac{d-3}{z} - \frac{\chi'}{2} \right)A_{t}' - \frac{2 q^2 A_{t} \phi^2}{z^2 \, f} \,, \\
0 & = \phi'' - \left( \frac{d-1}{z} - \frac{f'}{f} + \frac{\chi'}{2} \right)\phi' - \left( \frac{m^2}{z^2 \, f} -  \frac{q^2 A_{t}^2 e^{\chi}}{f^2} \right)\phi  \,,
\end{split}
\end{equation}
where a prime denotes a derivative with respect to $z$. The first two equations in Eq.~\eqref{eqsfchiAphi} come from the Einstein equation \eqref{eom3}\footnote{In fact, there are three nonzero Einstein equations, but only two of them are independent due to the Bianchi identity.}. The third and fourth are the Maxwell and scalar equations respectively.

There are two classes of the solutions for Eq.~\eqref{eqsfchiAphi}: i) $\phi=0$ and ii) $\phi\ne0$. They correspond to the normal phase and the superconductor phase respectively. This identification will be explained below Eq.~\eqref{asymphi1}.

For $\phi=0$, Eq.~\eqref{eqsfchiAphi} allows an analytic solution
\begin{equation}\label{trivialsol}
\begin{split}
&f(z) = 1 - \frac{z^d}{z_{h}^d} - \frac{d-2}{2(d-1)}\left( \frac{z^d}{z_{h}^d} - \frac{z^{2d-2}}{z_{h}^{2d-2}} \right) z_{h}^2 \, \mu^2 \,, \\
&A_t(z) = \mu\left(1-\frac{z^{d-2}}{z_{h}^{d-2}}\right)\,, \qquad \chi(z) = 0 \,.
\end{split}
\end{equation}

For $\phi\ne0$, there is no analytic solution for Eq.~\eqref{eqsfchiAphi} so we have to solve them numerically. For a concrete numerical computation, we will set $d=3$ from here.
In order to solve Eq. \eqref{eqsfchiAphi} we need total six initial or boundary conditions.

Because we are interested in the black brane solutions which have a regular event horizon at $z_h$, we set $f(z_h)=0$. The Hawking temperature then can be expressed as\footnote{In normal phase,
\begin{equation}\label{Hawkingnormal}
  T   = \frac{1}{z_{h}} \left( \frac{d}{4\pi} - \frac{(d-2)^2 z_{h}^2 \, \mu^2 }{8\pi (d-1)} \right) \,.
\end{equation}
}
\begin{equation}\label{HawkingT}
  T=-\left.\frac{e^{-\chi/2}f'}{4\pi}\right|_{z=z_h} \,.
\end{equation}
In addition, one must require $A_t(z_h) = 0$ in order for
$g_{\mu\nu}A^\mu A^\nu$ to be finite at the horizon. As the horizon is regular, all functions should have finite values and admit Taylor's expansions in terms of $z_h-z$ when $z\rightarrow z_h^-$.
Then, at the horizon, we have the following relations
\begin{align}\label{eqinit}
  \phi'=-\frac{4\phi m^2}{2m^2\phi^2+A_{t}^2-12} \,.
\end{align}
\begin{figure}[]
 \centering
     \subfigure[ $\chi(\tilde{z})$]
     {\includegraphics[width=7cm]{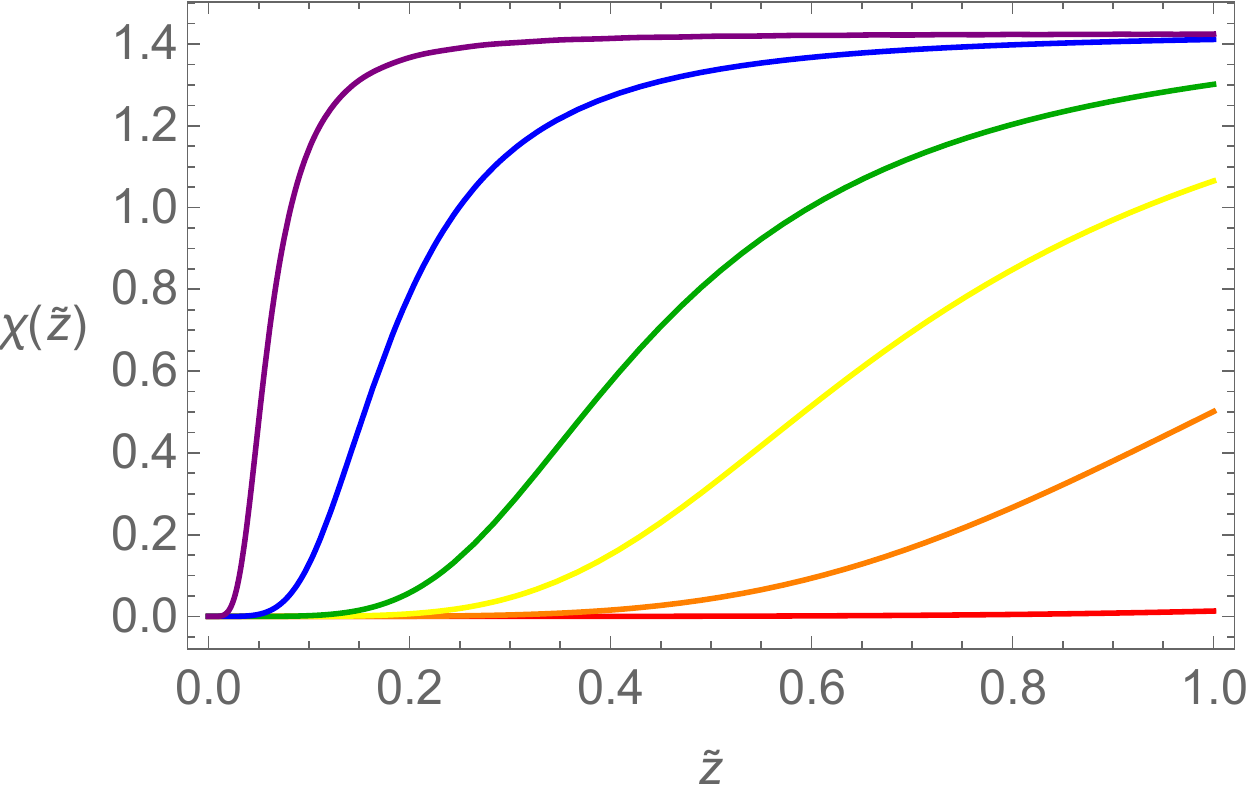} \label{}}
     \subfigure[$f(\tilde{z})$]
     {\includegraphics[width=7cm]{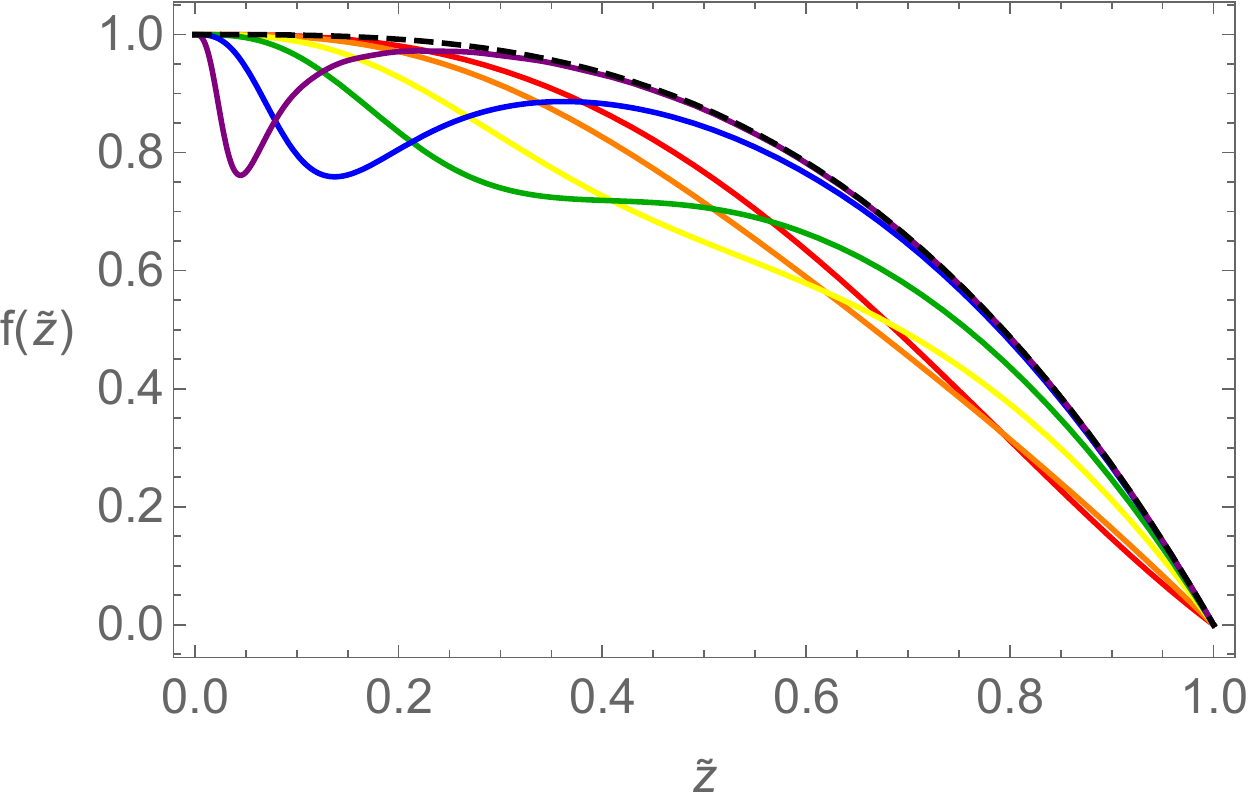} \label{m0PLOT2b}}
     \subfigure[ $A_{t}(\tilde{z})$]
     {\includegraphics[width=7cm]{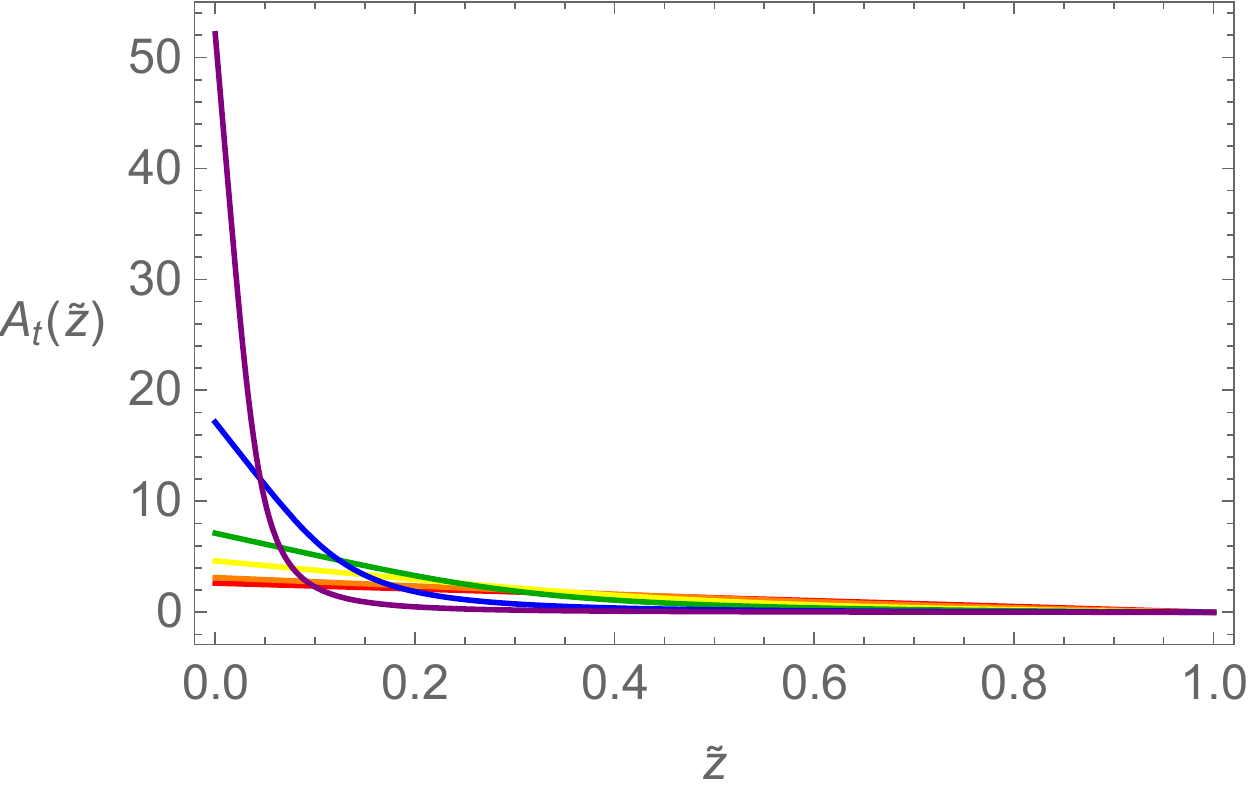} \label{}}
     \subfigure[ $\phi(\tilde{z})$]
     {\includegraphics[width=7cm]{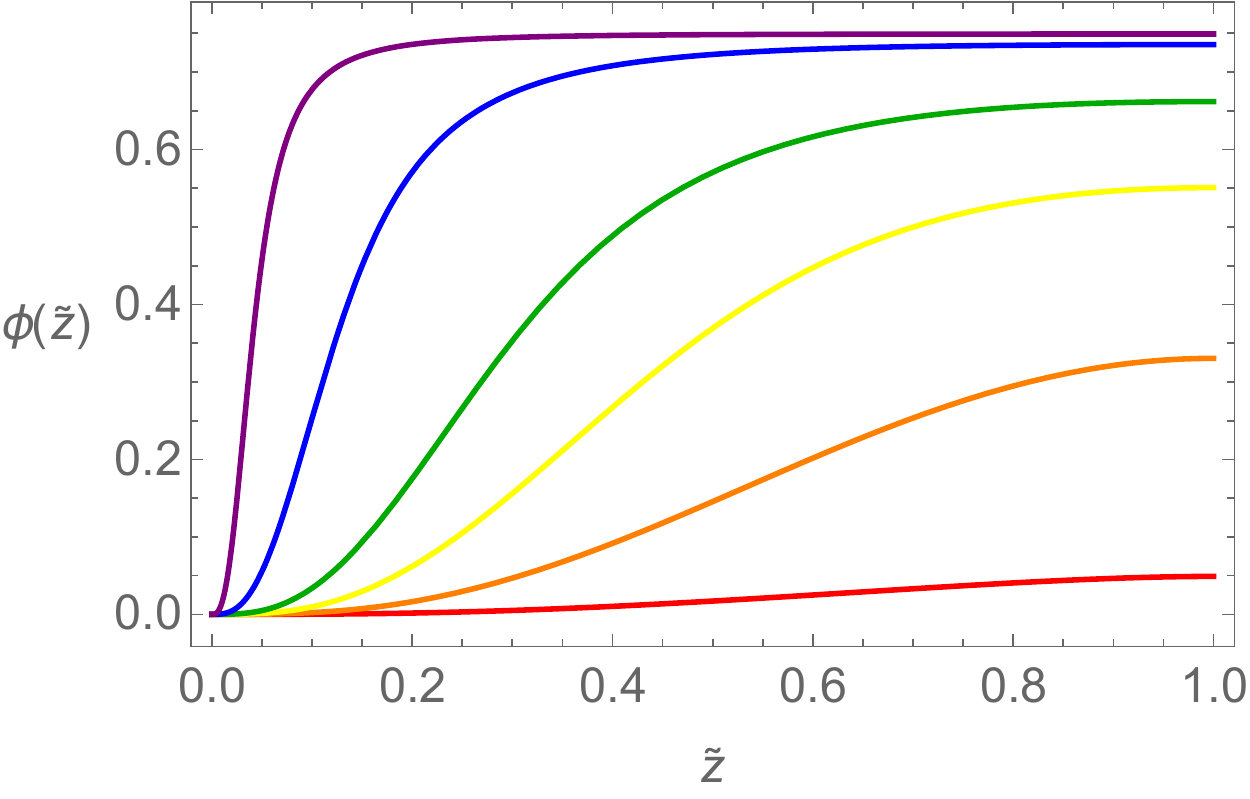} \label{}}
          \caption{Superconducting phase solutions for $(d, m^2, q)=(3, 0, 2.5)$.  The various colors represent different temperature. i.e. $T/T_{c}$ = 0.99, 0.86, 0.61, 0.41, 0.17, 0.05 (red, orange, yellow, green, blue, purple). In (b), the dashed black line is the AdS-Schwarzschild blackening factor: $f(\tilde{z}) = 1-\tilde{z}^3$. } \label{m0PLOT2}
\end{figure}

At the boundary $z\rightarrow0^+$, we want the space-time to asymptote to the AdS spacetime in Poincar\'{e} coordinate. Thus, we impose a boundary condition $\chi(0)=0$. It also turns out that  the asymptotic solutions for $\phi$ and $A_t$ are
\begin{equation}\label{asymphi1}
  \phi=\phi_+z^{-\Delta_+}+\phi_-z^{-\Delta_-}+\cdots\,, \qquad A_t=\mu-\rho z+\cdots \,,
\end{equation}
where $\Delta_\pm=-\frac32\pm\sqrt{4m^2+9}/2$. By the AdS/CFT correspondence the leading term $\phi_+$ corresponds to a source. We turn off the source, i.e. $\phi_+=0$ to describe a spontaneous symmetry breaking. Then, $\phi_-$ corresponds to condensate. Thus, if $\phi_- \ne 0$ the state is superconducting while  if $\phi_- = 0$ (or $\phi=0$) the state is normal.  In the asymptotic form of $A_t$, the leading term $\mu$ is chemical potential and the sub-leading coefficient $\rho$ is charge density.  If we consider the system in the grand canonical ensemble, we need to fix the chemical potential at the boundary.
Near the boundary, we have the following asymptotic behaviors for $f$ and $\chi$,
\begin{equation}\label{asymfchi}
  \chi=\chi_0+\chi_1z^{3+\sqrt{4m^2+9}}+\cdots\,, \qquad f=1+f_0z^3+\cdots\,,
\end{equation}
for some constants $\chi_0$, $\chi_1$ and $f_0$.
For numerical analysis, we used the shooting method and pseudo-spectral method independently and have cross-checked our results.
For example, we show the numerical solutions of \eqref{eqsfchiAphi} for  $(d, m^2, q)=(3, 0, 2.5)$ in Fig \ref{m0PLOT2}, where $\tilde{z} := z/z_h$.

%

%
%
%
%
%

\section{Complexity of formation}

The complexity of formation by the finite temperature was first investigated by Ref. \cite{Chapman:2016hwi}. It is the complexity between the  AdS-Schwarzschild black brane and pure AdS space-time for a very special maximum space-like slice. In this section, we consider the complexity of formation for the holographic superconductor in both normal phase and superconducting phase. The complexity of formation in normal phase also has been reported in \cite{Carmi:2017jqz}. Here, we revisit it by a little different analytic method, and our results agree with \cite{Carmi:2017jqz}.

Before starting, let us introduce a new notation for the CV conjecture, different from  \eqref{CV}.
\begin{equation}\label{CV1}
  \mathcal{V}=\max_{\partial \Sigma=t_L\cup t_R} V(\Sigma)  \,,
\end{equation}
where we set $G_N =\ell =1 $ for convenience.


\subsection{Normal phase}
Following \cite{Chapman:2016hwi}, we consider the complexity of the thermal state defined on the time slice at $t_L = t_R = 0$. Because of the symmetry, the maximal volume is given by the $t = 0$ slice in bulk, i.e., the straight line connecting the two boundaries through the outer bifurcation horizon in the Penrose diagram shown in Fig. \ref{Fig1}.

The volume integral then simplifies to:
\begin{figure}
  \centering
  \includegraphics[width=\textwidth]{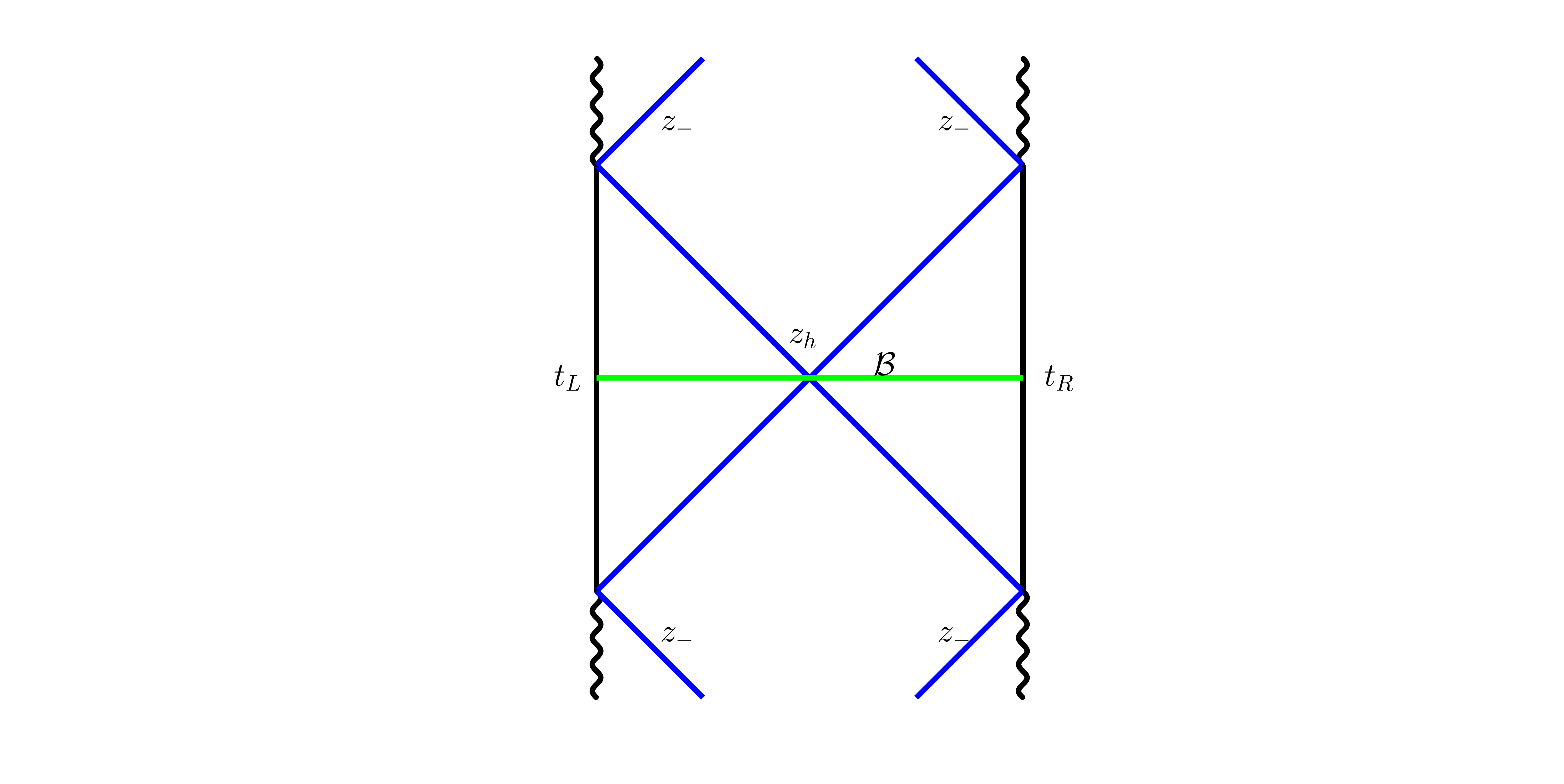}
  \caption{The maximal volume slice $\mathcal{B}$ connecting the two boundaries at $t_L = t_R = 0$ in one patch of the AdS-RN black hole.}\label{Fig1}
\end{figure}
\begin{equation}\label{CVRN1}
  \mathcal{V} = 2 \Omega_{d-1} \int^{z_h}_{z_{m}}\frac{\td z}{z^d \sqrt{f(z)}} \,,
\end{equation}
where $z_{m}$ is the cut-off near the boundary and $\Omega_{d-1}$ is the dimensionless area of the spatial geometry when $z$ and $t$ are fixed. With the general asymptotic behavior \eqref{asymfchi}, we find that $\mathcal{V}$ is infinite when $z_{m}\rightarrow0$. To evaluate the complexity of formation coming from finite temperature and chemical potential, we will subtract from this integral, the corresponding contribution from (two copies of) the vacuum AdS background,
\begin{equation}\label{CVAdS1}
\begin{split}
  \mathcal{V}_0 &= 2\Omega_{d-1}\int^{\infty}_{z_{m}}\frac{\td z}{z^d} \\
  &= 2\Omega_{d-1} \left(\int^{z_h}_{z_{m}} + \int^{\infty}_{z_h}\right) \frac{\td z}{z^d} = 2\Omega_{d-1} \left(  \int^{z_h}_{z_{m}}\frac{\td z}{z^d} + \frac{1}{d-1}\frac{1}{z^{d-1}_h}  \right).
  \end{split}
\end{equation}
Now one can find that,
\begin{equation}\label{deltaCVRN}
\begin{split}
  \Delta\mathcal{V} &= \mathcal{V}-\mathcal{V}_0 \\
                               &=  2\Omega_{d-1} \left[ \int^{z_h}_{z_{m}}\frac{\td z}{z^d}\left(\frac1{\sqrt{f(z)}}-1\right) - \frac{1}{d-1}\frac{1}{z^{d-1}_h} \right] \\
                               &=  \frac{2\Omega_{d-1}}{z^{d-1}_{h}} \left[ \int^{1}_{\tilde{z}_{m}}\frac{\td \tilde{z}}{\tilde{z}^d}\left(\frac1{\sqrt{f(\tilde{z})}}-1\right) - \frac{1}{d-1} \right] \,, \\
\end{split}
\end{equation}
where
\begin{equation} \label{ztilde}
\tilde{z} :=z/z_{h} \,, \qquad \tilde{z}_{m} :=z_{m}/z_{h} \,.
\end{equation}

The emblackening factor $f(z)$ in Eq.~\eqref{trivialsol} and the temperature\eqref{Hawkingnormal} read
\begin{align}
     f(\tilde{z}) &= 1 - \tilde{z}^d - \frac{d-2}{2(d-1)}\left( \tilde{z}^d - \tilde{z}^{2d-2} \right) \tilde{\mu}^2 \,,\label{tildeFormula1} \\
                 T  &=  \frac{1}{z_{h}} \left( \frac{d}{4\pi} - \frac{(d-2)^2}{8\pi (d-1)}\tilde{\mu}^2 \right) =: \frac{1}{z_{h}} \tilde{T} \,, \label{tildeFormula2}
\end{align}
where we define $\tilde{T} :=T z_{h}$ and $\tilde{\mu} :=\mu z_{h}$.

The dimensionless variables with tilde denote the variables scaled by $z_{h}$ and convenient for numerical analysis. However, for physical interpretation it is better to choose the chemical potential as our scale. Thus we also define the dimensionless variables with bar to denote the variables scaled by the chemical potential. For example,
\begin{align}
     \Delta\bar{\mathcal{V}}_d^{\mathrm{n}}(\bar{T})  &:=  \frac{\Delta\mathcal{V}}{\mu^{d-1}\Omega_{d-1}} = \frac{2}{\tilde{\mu}^{d-1}} \left[ \int^{1}_{\tilde{z}_{m}}\frac{\td \tilde{z}}{\tilde{z}^d}\left(\frac1{\sqrt{f(\tilde{z})}}-1\right) - \frac{1}{d-1} \right] \,, \label{barFormula1} \\
                               \bar{T}  &:=  \frac{T}{\mu} = \frac{\tilde{T}}{\tilde{\mu}} =  \frac{1}{\tilde{\mu}} \left( \frac{d}{4\pi} - \frac{(d-2)^2}{8\pi (d-1)}\tilde{\mu}^2 \right) \,, \label{barFormula2}
\end{align}
where the indices `n' and '$d$' of $\Delta\bar{\mathcal{V}}^{\mathrm{n}}_d$ means `n'ormal state in `$d$' spatial dimension.   In the definition of $\Delta\bar{\mathcal{V}}^{\mathrm{n}}_d$ we included the trivial volume factor $\Omega_{d-1}$.
The complexity of formation $\Delta\bar{\mathcal{V}}^{\mathrm{n}}_d$  is a function of $\bar{T}$, because $\tilde{\mu}$ is a function of $\bar{T}$ for a given dimension $d$ by the relation \eqref{barFormula2}. $\tilde{z}_m$ will be taken to be zero at the end of the day.

\begin{figure}[]
 \centering
     {\includegraphics[width=8cm]{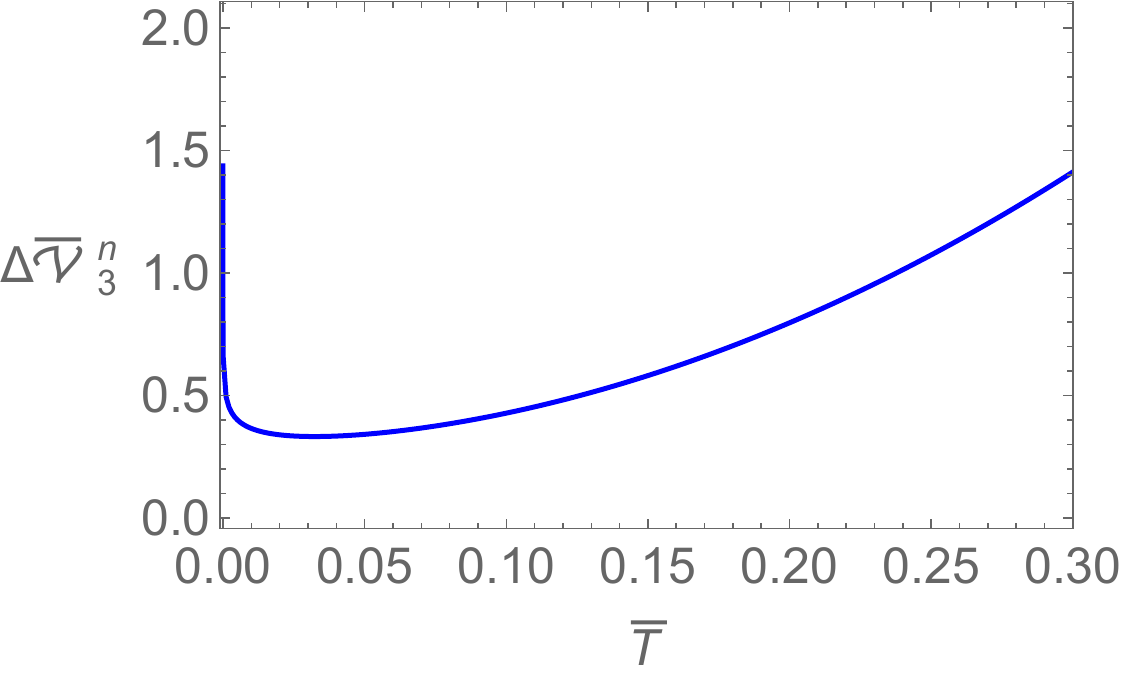} \label{}}
          \caption{ The complexity of formation in normal phase. The numerical plot of \eqref{barFormula1} with $d=3$.} \label{FIGnormal}
\end{figure}

Eq.~\eqref{barFormula1} does not allow an analytic expression in general,
so we perform the integration numerically. The result is shown in Fig. \ref{FIGnormal}.
As the temperature goes to zero and goes to infinity, the complexity of formation diverges in both cases. There is a particular temperature in the middle where the complexity of formation is minimum. This has been first observed in \cite{Carmi:2017jqz}.

To analyze the divergence behavior at low and high temperature, we consider the expansion in terms of $\bar{T}$.
First, in these limits the chemical potential reads from~\eqref{barFormula2}
\begin{align}
\tilde{\mu} &= \frac{d}{4\pi \bar{T}}   - \frac{(d-2)^2 d^2}{128 \pi^3 (d-1) \bar{T}^3}    \, + \, \mathcal{O}(\bar{T}^{-5}) \,, \qquad\qquad\quad (\bar{T} \gg 1) \,, \label{mulimit1} \\
\tilde{\mu} &= \sqrt{\frac{2d(d-1)}{(d-2)^2}} \, - \, \frac{4\pi(d-1)}{(d-2)^2} \bar{T}  \, + \, \mathcal{O}\left(\bar{T}^{2}\right) \,. \qquad\quad\,\,\, (\bar{T} \ll 1) \,. \label{mulimit2}
\end{align}
so the emblackening factor in \eqref{tildeFormula1} behaves as follows:
\begin{align}
f(\tilde{z}) &= 1-\tilde{z}^{d} - \frac{(d-2)d^2}{32 \pi^2 (d-1)}\frac{\tilde{z}^d - \tilde{z}^{2d-2}}{\bar{T}^2}     + \mathcal{O}(\bar{T}^{-4}) \,,  \,\,\,\,  \qquad\qquad\qquad  (\bar{T} \gg 1) \,,     \label{BFlimit1} \\
f(\tilde{z}) &= 1-\frac{2(d-1)}{d-2}\tilde{z}^d +\frac{d}{d-2}\tilde{z}^{2d-2}   + \sqrt{\frac{32d(d-1)\pi^2}{(d-2)^4}}(\tilde{z}^d-\tilde{z}^{2d-2}) \bar{T}      + \mathcal{O}(\bar{T}^2)     \qquad\,  (\bar{T} \ll 1) \,. \label{BFlimit2}
\end{align}

\begin{figure}[]
 \centering
     \subfigure[High temperature: the red line is \eqref{HighTbehavior2}.]
     {\includegraphics[width=7.33 cm]{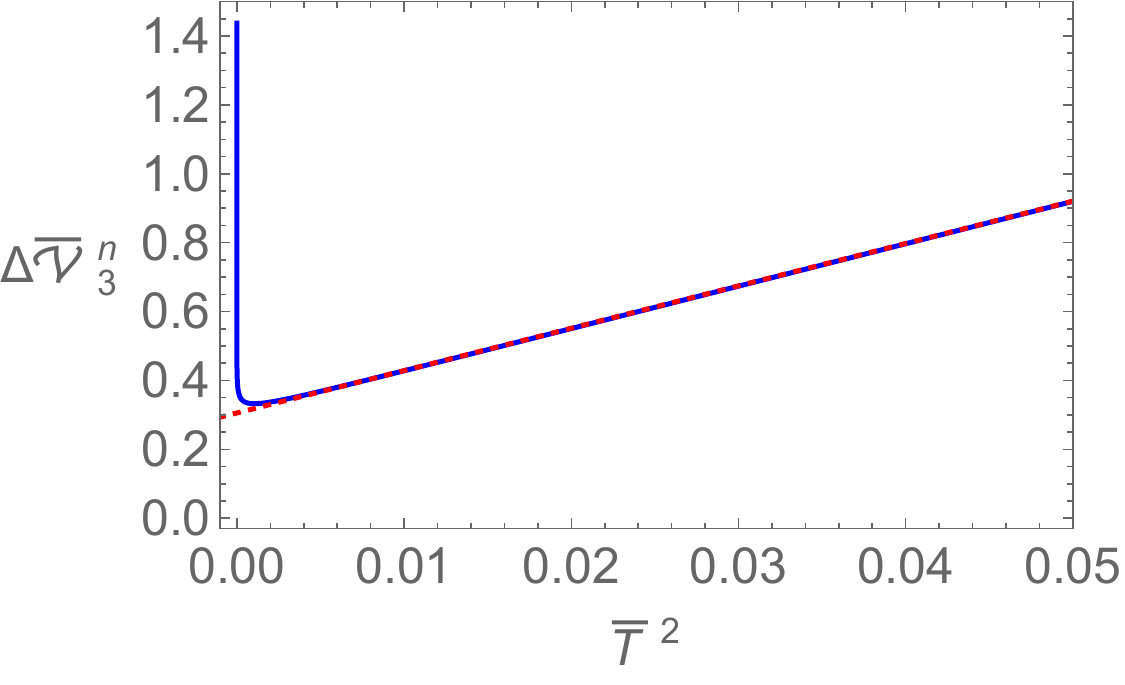} \label{FIGnormal1a}} \ \
     \subfigure[Low temperature: the red line is \eqref{LowTvol6}]
     {\includegraphics[width=7.1 cm]{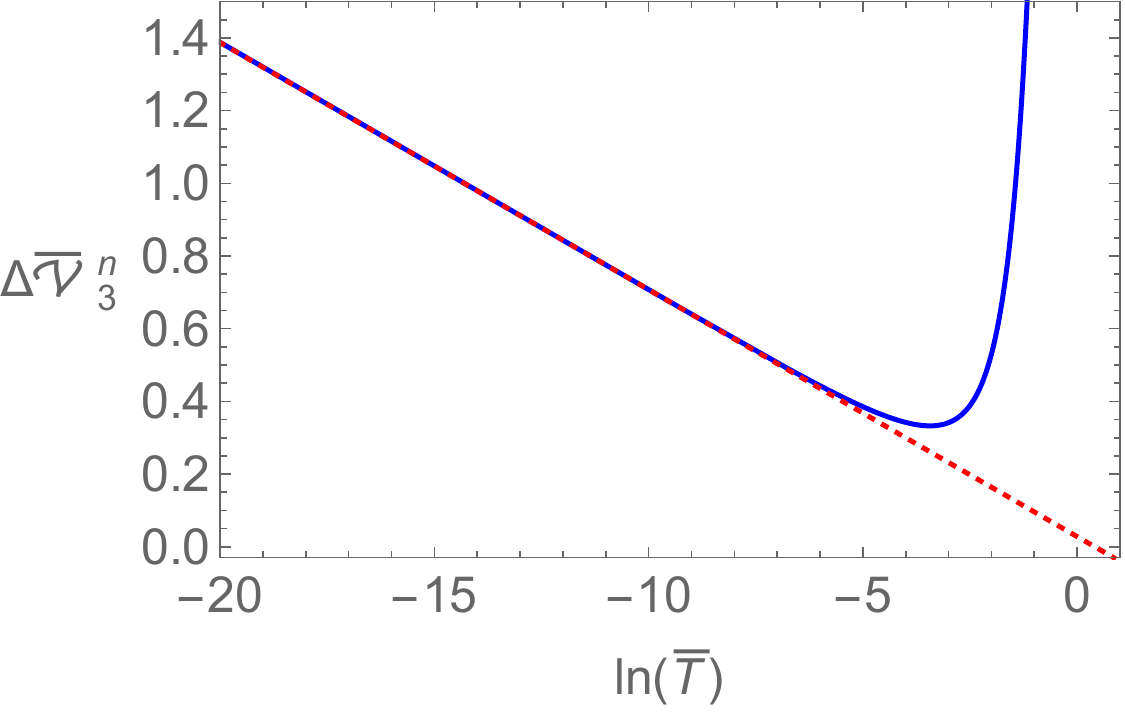} \label{FIGnormal1b}}
          \caption{ The low and high temperature behavior of the complexity of formation in normal phase. The blue curve is the numerical plot of \eqref{barFormula1} with $d=3$.} \label{FIGnormal1}
\end{figure}

In the high temperature limit($\bar{T} \gg 1$), by using \eqref{BFlimit1} and \eqref{mulimit1}, \eqref{barFormula1} becomes
\begin{equation}\label{HighTbehavior2}
\begin{split}
\Delta\bar{\mathcal{V}}^{\mathrm{n}}_d (\bar{T})
=& \frac{(4\pi)^{\frac{2d-1}{2}}}{d^d} \frac{\Gamma\left(\frac{1-d}{d}\right)}{\Gamma\left(\frac{2-d}{2d} \right)} \, \bar{T}^{d-1}  \\
&+   \frac{(4\pi)^\frac{2d-5}{2}(d-2)}{2d^{d-1}(d-1)} \left(  \frac{(d-1)^{2}\Gamma\left(\frac{1-d}{d}\right)}{\Gamma\left(\frac{2-d}{2d}\right)}   +   \frac{d \Gamma\left(\frac{d-1}{d}\right)}{\Gamma\left(\frac{d-2}{2d}\right)}\right) \bar{T}^{d-3}  \,+\,    \mathcal{O}(\bar{T}^{d-5})     \,.
\end{split}
\end{equation}
For $d=3$ case, Eq. \eqref{HighTbehavior2} becomes $\Delta\bar{\mathcal{V}}^{\mathrm{n}}_3 = 12.301 \bar{T}^2 + 0.305$, which is confirmed in our numerical computation. See the red line in Fig \ref{FIGnormal1a}.

In the low temperature limit ($\bar{T} \ll 1$), the emblackening factor $f(\tilde{z})$ in \eqref{BFlimit2}  behaves near the horizon($\tilde{z}=1$) as follows:
\begin{equation}\label{nearhorizonLowT}
\begin{split}
f(\tilde{z}) &= d(d-1)(\tilde{z}-1)^2    \\ &+   \left( -\sqrt{\frac{ 72 d^3 (d-1) \pi^2  }{ (d-2)^2  }}(\tilde{z}-1)^2   -  \sqrt{\frac{ 32d(d-1)\pi^2 }{ (d-2)^2 }}(\tilde{z}-1)  \right)\bar{T}
 +\, \mathcal{O}(\bar{T}^2,(\tilde{z}-1)^3)    \,.
\end{split}
\end{equation}
Since $f(\tilde{z})$ goes to zero quadratically near the horizon, the integral \eqref{barFormula1} diverges logarithmically at the zero temperature limit.

To deal with this divergence separately we introduce $\tilde{a}$ such that $\tilde{z}_{m} <\tilde{a} <1$ and divide the integral range as
\begin{equation}\label{LowTvol1}
\begin{split}
     \Delta\bar{\mathcal{V}}^{\mathrm{n}}_d(\bar{T})  &= \frac{2}{\tilde{\mu}^{d-1}}  \left[\left(\int^{\tilde{a}}_{\tilde{z}_{m}} + \int^{1}_{\tilde{a}}\right)\frac{\td \tilde{z}}{\tilde{z}^d}\left(\frac1{\sqrt{f(\tilde{z})}}-1\right) - \frac{1}{d-1} \right]  \,, \\
                                           &= \frac{2}{\tilde{\mu}^{d-1}}  \left[\int^{1}_{\tilde{a}}\frac{\td \tilde{z}}{\tilde{z}^d} \frac1{\sqrt{f(\tilde{z})}}    \,+\,   \mathcal{F}_{1}(\tilde{a},d,\bar{T})         \right]  \,,
\end{split}
\end{equation}
where $\mathcal{F}_{1}(\tilde{a},d,\bar{T})$ is a finite piece,
\begin{equation}\label{finite1}
        \mathcal{F}_{1}(\tilde{a},d,\bar{T})  :=  \int^{\tilde{a}}_{\tilde{z}_{m}}\frac{\td \tilde{z}}{\tilde{z}^d} \left(\frac1{\sqrt{f(\tilde{z})}} - 1 \right) - \frac{\tilde{a}^{1-d}}{d-1}  \,.
\end{equation}
To have a better understanding of the singular term in \eqref{LowTvol1}, we define the followings:
\begin{align}
       h(\tilde{z},\bar{T})  &:= \frac{\tilde{z}^{2d} f(\tilde{z})}{(\tilde{z}-1)(\tilde{z}-1-\Gamma(\bar{T}))}   \,. \label{gammah2}  \\
       \Gamma(\bar{T}) &= \gamma \bar{T}  \,+\, \mathcal{O}(\bar{T}^2) \,, \qquad \gamma := \sqrt{\frac{ 32 \pi^2 }{ d(d-1)(d-2)^2 }}  \,,\label{gammah22}
\end{align}
where $\Gamma(\bar{T})$ is chosen such that  $h(\tilde{z}, \bar{T}) $ is regular and well-defined at $\tilde{z}=1$ and  $\bar{T}=0$, i.e.
\begin{equation} \label{cond0}
h(1,0) := \lim_{\bar{T} \rightarrow 0}\lim_{\tilde{z} \rightarrow 1}  h(\tilde{z}, \bar{T}) = \lim_{\tilde{z} \rightarrow 1} \lim_{\bar{T} \rightarrow 0}  h(\tilde{z}, \bar{T})   \,.
\end{equation}

Let us explain how to determine \eqref{gammah22}.
The first condition in \eqref{cond0} may be rephrased as
\begin{equation}
       h(\tilde{z} = 1,\bar{T} \ll 1)  = \frac{  \sqrt{\frac{ 32d(d-1)\pi^2 }{ (d-2)^2 }}  \bar{T}   + \mathcal{O}({\bar{T}^2})   }{\Gamma(\bar{T}\ll1)}  \,,   \label{gammah4}
\end{equation}
where \eqref{gammah2} and  \eqref{nearhorizonLowT} are used. To have a regular $h(1,0)$ we need to require
\begin{align}
       \Gamma(\bar{T}\ll1) := \gamma \bar{T}  \,+\, \mathcal{O}(\bar{T}^2) \,.   \label{Gammagamma}
\end{align}
The second condition of \eqref{cond0} implies
\begin{align}
       h(1,0)  = \frac{  \sqrt{\frac{ 32d(d-1)\pi^2 }{ (d-2)^2 }}     }{\gamma}   = d(d-1)  \,,  \label{constant1}
\end{align}
which yields
\begin{align}
        \gamma = \sqrt{\frac{ 32 \pi^2 }{ d(d-1)(d-2)^2 }}      \,.\label{h0h1gamma}   		
\end{align}

By using  \eqref{gammah2}, we rewrite \eqref{LowTvol1} as follows:
\begin{equation}\label{LowTvol2}
\begin{split}
     \Delta\bar{\mathcal{V}}^{\mathrm{n}}_d(\bar{T}) &=  \frac{2}{\tilde{\mu}^{d-1}}  \left[\int^{1}_{\tilde{a}}\td \tilde{z} \frac1{\sqrt{(\tilde{z}-1)(\tilde{z}-1-\Gamma(\bar{T}))h(\tilde{z},\bar{T})}}    \,+\,   \mathcal{F}_{1}(\tilde{a},d,\bar{T})         \right] \,.   \\
    &  =   \mathcal{S}(\tilde{a},d,\bar{T})     +   \frac{2}{\tilde{\mu}^{d-1}}  \left[\mathcal{F}_{2}(\tilde{a},d,\bar{T})  + \mathcal{F}_{1}(\tilde{a},d,\bar{T})  \right]  \,,
\end{split}
\end{equation}
where  $\mathcal{S}(\tilde{a},d,\bar{T})$ is a singular term
\begin{align}\label{LowTvol3}
\begin{split}
 \mathcal{S}(\tilde{a},d,\bar{T})  &:=  \frac{2}{\tilde{\mu}^{d-1}}  \int^{1}_{\tilde{a}}\td \tilde{z} \frac1{\sqrt{(\tilde{z}-1)(\tilde{z}-1-\Gamma(\bar{T}))h(1,\bar{T})}}  \\
     &  =   \frac{2}{\tilde{\mu}^{d-1}}  \frac{1}{\sqrt{h(1,\bar{T})}} \ln\left( \frac{\Gamma(\bar{T})}{\Gamma(\bar{T})-2(\tilde{a}-1)-2\sqrt{(\tilde{a}-1)(\tilde{a}-1-\Gamma(\bar{T}))}} \right)    \,,
\end{split}
\end{align}
and $\mathcal{F}_{2}(\tilde{a},d,\bar{T})$ is another finite term defined as
\begin{equation} \label{finite21}
\mathcal{F}_{2}(\tilde{a},d,\bar{T})   :=  \int^{1}_{\tilde{a}}\td \tilde{z} \left(\frac1{\sqrt{(\tilde{z}-1)(\tilde{z}-1-\Gamma(\bar{T}))h(\tilde{z},\bar{T})}}  - \frac1{\sqrt{(\tilde{z}-1)(\tilde{z}-1-\Gamma(\bar{T}))h(1,\bar{T})}}  \right)   \,,
\end{equation}

With the formulas \eqref{LowTvol3}-\eqref{finite21} and the expansion for $\tilde{\mu}$ in \eqref{mulimit2}, the complexity of formation \eqref{LowTvol2} can be expanded in terms of $\bar{T}$ as
\begin{equation}\label{LowTvol6}
\begin{split}
     \Delta\bar{\mathcal{V}}^{\mathrm{n}}_d(\bar{T}) =    -\sqrt{\frac{2^{3-d}}{d^d (d-1)^d (d-2)^{2(1-d)}}} \, \ln(\bar{T})   \,+\,   \mathcal{F}_{3}(\tilde{a},d)    \,+\,  \mathcal{O}\left(\bar{T}\right) \,,
\end{split}
\end{equation}
where
\begin{equation}\label{finite3}
\begin{split}
    \mathcal{F}_{3}(\tilde{a},d) &:=   \frac{2^{\frac{3-d}{2}}d^{\frac{1-d}{2}}(d-1)^{\frac{1-d}{2}}}{(d-2)^{1-d}} \left[\frac{1}{\sqrt{d(d-1)}}\ln\left(   \frac{    4(1-\tilde{a})   }{   \gamma    }   \right)  +  \mathcal{F}_{2}(\tilde{a},d,0)  + \mathcal{F}_{1}(\tilde{a},d,0) \right]   \,.  \\
\end{split}
\end{equation}

For example, if $d=3$,
\begin{align}
        \mathcal{F}_{1}(\tilde{a},3,0) & =  \frac{(\tilde{a}-1)\sqrt{1+2\tilde{a}+3\tilde{a}^2}}{2\tilde{a}^2} - \frac{1}{\sqrt{6}} \ln\left( \frac{(2+\sqrt{6})(1-\tilde{a})}{2+4\tilde{a}+\sqrt{6(1+2\tilde{a}+3\tilde{a}^2)}} \right)      \,, \label{finited3val21}  \\
         \mathcal{F}_{2}(\tilde{a},3, 0)   &=  \frac{(1-\tilde{a})\sqrt{1+2\tilde{a}+3\tilde{a}^2}}{2\tilde{a}^2} + \frac{1}{\sqrt{6}} \ln\left( \frac{12}{2+4\tilde{a}+\sqrt{6(1+2\tilde{a}+3\tilde{a}^2)}} \right)    \,.
\end{align}
with $\tilde{z}_{m}=0$ and
\begin{align}\label{}
       \mathcal{F}_{3}(\tilde{a},3) & =  \frac{1}{6\sqrt{6}} \ln\left( \frac{12\sqrt{3}}{\pi(2+\sqrt{6})} \right) \approx 0.0269 \,. \label{finited3val3}
\end{align}
Note that $\mathcal{F}_{3}(\tilde{a},3)$ is independent of the arbitrary choice of $\tilde{a}$. Finally,
\begin{equation}
\Delta\bar{\mathcal{V}}^{\mathrm{n}}_3(\bar{T})  = -0.068\ln(\bar{T})+0.0269 \,,
\end{equation}
which is confirmed in our numerical computation. See the red line in Fig \ref{FIGnormal1b}.


\subsection{Superconducting phase}
In superconducting phase, the non-trivial complex scaler field plays a role and the low temperature behavior will be different from the RN-AdS black holes. At low temperature below some critical temperature $T_c$, the superconductor state with $\phi \ne 0$ has the lower free energy than a normal state with $\phi =0$ so it becomes the  ground state.  For example,  for $q=1, m^2=-2$,  the dimensionless critical temperature $T_c/\mu\simeq0.0208$.  In this section, we study the complexity of formation in the superconducting phase.

\begin{figure}
  \centering
  \includegraphics[width=.8\textwidth]{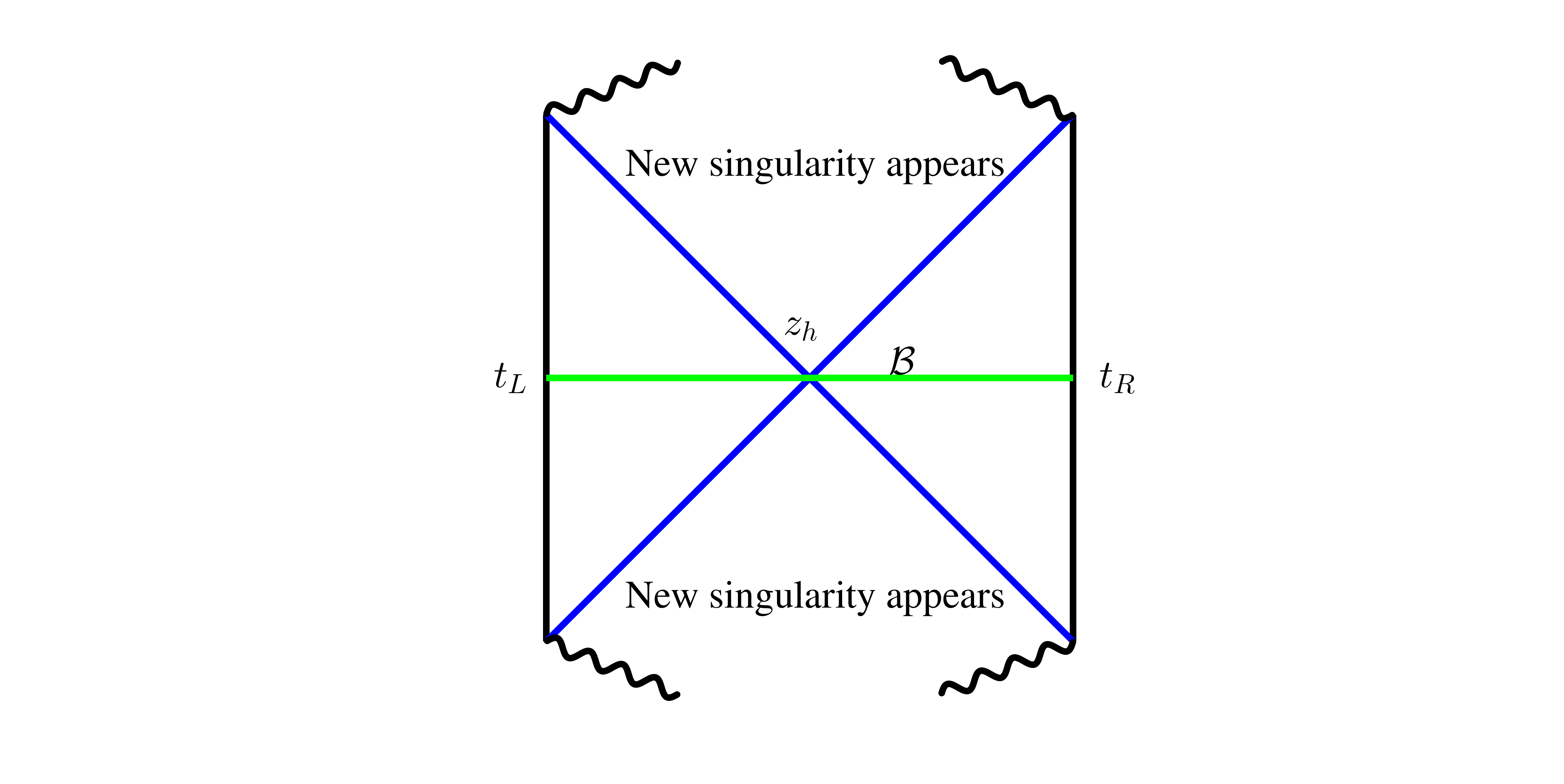}
  \caption{The schematic Penrose diagram in the connected region which contains an Einstein-Rosen bridge to connect two AdS boundaries. The maximal volume slice $\mathcal{B}$ for $t_L = t_R = 0$  is the same as AdS RN black brane. When complex scalar field appears, the Cauchy horizon (inner horizon) will be taken place by a singularity. }\label{FigP2}
\end{figure}
For $\phi \ne 0$, the geometry outside of the event horizon is still regular, so the causal structure outside of $z_h$ is similar to the RN black hole. However, in this case the inner horizon disappears and a new singularity appears.
Due to the existence of the Cauchy horizon in the RN-AdS black hole, the complete description of the inner geometry of the charged black hole after the scalar hair appears is subtle.
Some analyses show that the Cauchy horizon will become a ``weak singularity'' if the scalar field is infinitesimal~\cite{Ori:1991aa,Burko:1997aa}. As the scalar field increases, it is not clear how such a new ``weak singularity'' will evolve.
In Fig. \ref{FigP2}, we show a schematic Penrose diagram in the connected region which contains an Einstein-Rosen bridge between two AdS boundaries. Though the structure of the singularity is not clear, the computation of the CV conjecture is not affected by the detailed structure of singularity. However, for the CA conjecture, as the WdW patch touches  the singularity in general, we have to know the detailed properties of the  singularity. Partly because of this subtlety, in this paper, we only focus on the  CV conjecture.

The maximal volume slice $\mathcal{B}$ for $t_L = t_R = 0$  is the same as the RN black hole. Thus, the complexity for $t_L=t_R=0$ in superconducting phase is still given by \eqref{CVRN1}.
Taking into account the general asymptotic solution for metric shown in  \eqref{asymfchi} at the boundary $z=0$, we find that the $\mathcal{V}$ diverges but the complexity of formation $\Delta \mathcal{V}$  in  \eqref{deltaCVRN} and $ \Delta\bar{\mathcal{V}}^\mathrm{n}_d$ in \eqref{barFormula1} are still valid also for the superconducting phase.
\begin{equation} \label{super11}
     \Delta\bar{\mathcal{V}}^\mathrm{s}_d(\bar{T}) :=  \frac{\Delta\mathcal{V}}{\mu^{d-1}\Omega_{d-1}} = \frac{2}{\tilde{\mu}^{d-1}} \left[ \int^{1}_{\tilde{z}_{m}}\frac{\td \tilde{z}}{\tilde{z}^d}\left(\frac1{\sqrt{f(\tilde{z})}}-1\right) - \frac{1}{d-1} \right] \,,
     \end{equation}
where we changed the superscript `n' to `s' to make clear we are considering the superconducting phase. Contrary to normal phase, in this case, there is no analytic formula for the emblackening factor $f(\tilde{z})$ is available so we resort to the numerical methods\footnote{The temperature \eqref{barFormula2} is only valid for normal phase. For superconducting phase, we should come back to the original definition \eqref{HawkingT}.}.

Note that the formula \eqref{super11} is valid for both normal phase and superconducting phase, so we will use the subscript `n' and `s' only we want to emphasize which phase we are looking at.

\begin{figure}
  \centering
     \subfigure[$\Delta\bar{\mathcal{V}}_3$ and $\delta\bar{\mathcal{V}}_3$ vs $T/T_c$]
  {\includegraphics[width=6.2cm]{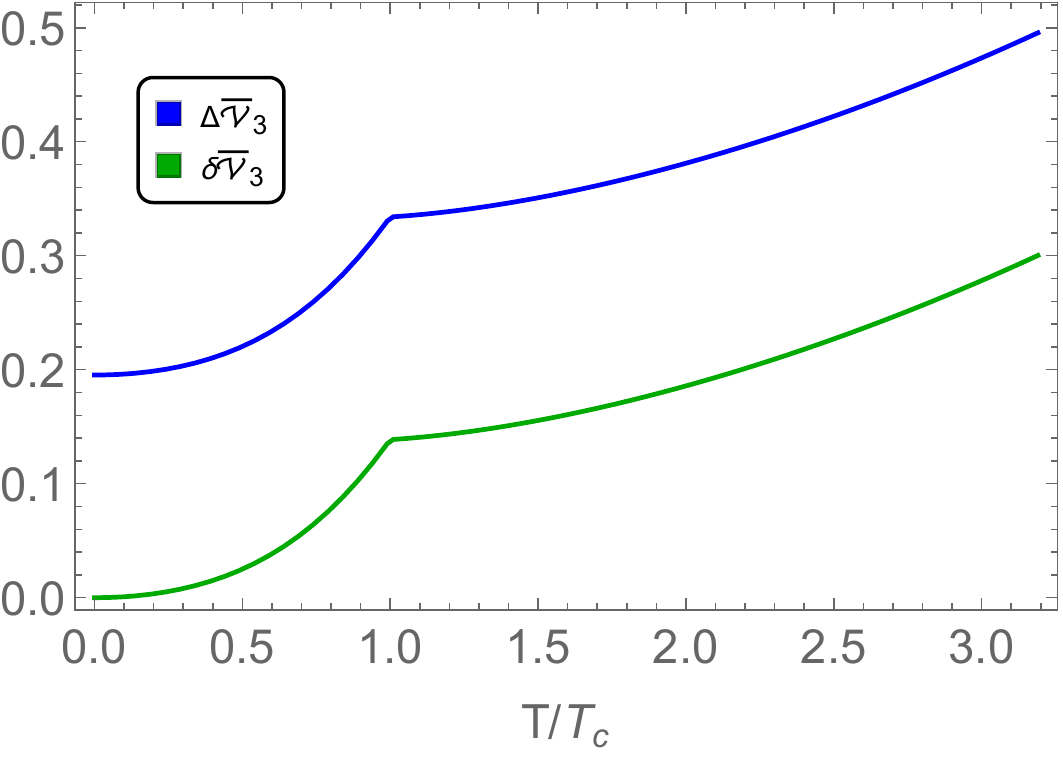}\label{Fig4a}} \ \  \
     \subfigure[$\ln(\delta\bar{\mathcal{V}}_3^\mathrm{s})$ vs. $\ln(T/T_c)$]
  {\includegraphics[width=7.7cm]{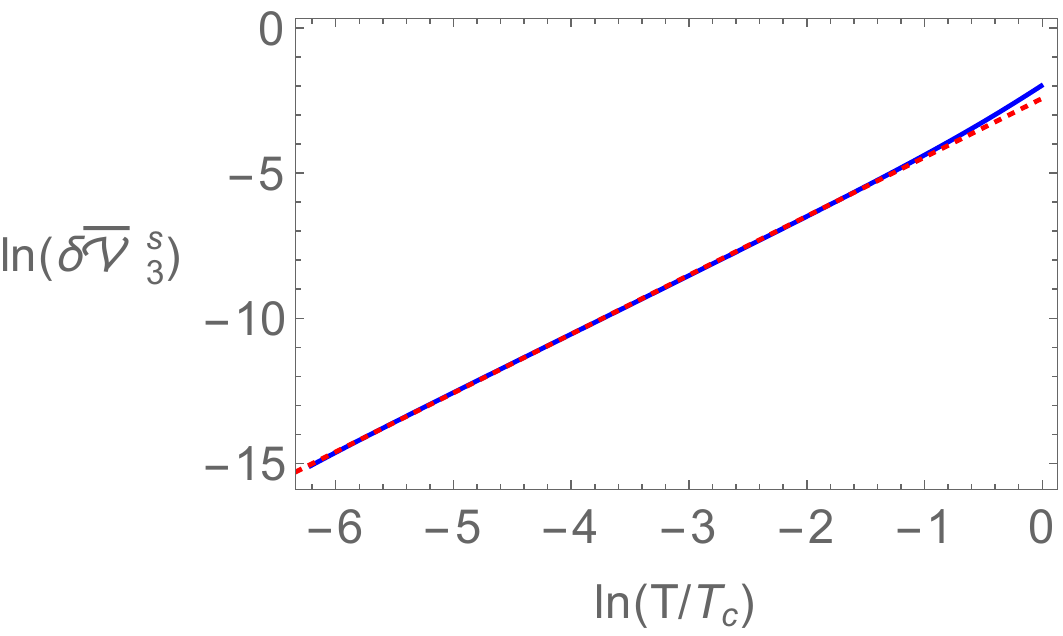}\label{Fig4b}}
  \caption{The complexity of formation and thermal complexity of the holographic superconductor with { $m^2=0$ and $q= 2.5$. $T_c/\mu\simeq 0.039 $ is the critical temperature. }}
  \end{figure}

\subsubsection{Complex scalar mass: $m^2=0$} \label{m20case}

Fig.\ref{Fig4a} is the numerical plots of the complexity formation for the system with $m^2=0, q=2.5$. The horizontal axis is $T/T_c$ where $T_c$ is the critical temperature of the phase transition. For $T/T_c >1$ the system is in normal phase and the complexity is the same as Fig.\ref{FIGnormal}. At $T/T_c=1$ there is a kink and the transition is not smooth.  For $T/T_c <1$ the complexity monotonically decreases contrary to the normal phase. The complexity in superconducting phase is smaller than normal phase similar to the free energy. {We also have analyzed  many cases with different masses $m^2$ and charges $q$, and found a universal property: {\it the superconducting phase always has the smaller complexity than the unstable normal phase below the critical temperature.} It seems that the complexity may play a role of the free energy and indicates the phase transition\footnote{In \cite{Ali:2018aon} it is also shown that complexity serves as an order parameter in a different system (topological system).  The complexity near superconducting transition point has been studied also in \cite{Momeni:2016ekm} in the context of ``holographic complexity'' proposed by \cite{Alishahiha:2015rta}, which is different from CV conjecture.
\cite{Flory:2017ftd} also dealt with the behaviour of CV complexity in a holographic model with a superconductor-like phase transition, and found that the complexity decreases in the condensed phase.}. It will be  interesting to investigate if this property is indeed universal, and understand why, if it is. }

At zero temperature the complexity of formation $\Delta\bar{\mathcal{V}}^\mathrm{s}_d(\bar{T}=0)$  is nonzero and it
is the complexity of the zero temperature superconductor state from the AdS vacuum state. We may also define the complexity of formation from  the zero temperature superconductor not from the AdS vacuum state:
\begin{align}
  \delta\bar{\mathcal{V}}^s_d({\bar{T}}) &:=  \bar{\mathcal{V}}-\bar{\mathcal{V}}_{\text{zero temperature superconductor}}  \\
 &= \Delta\bar{\mathcal{V}}_d^s(\bar{T}) - \Delta\bar{\mathcal{V}}_d^s(0) \,. \label{deltaVs}
\end{align}
Indeed, this new complexity of formation has a more natural and phenomenological interpretation because the ground state of the superconductor is zero temperature superconductor not a vacuum.
To distinguish this new complexity of formation, we will call it ``thermal complexity'' from here.

In Fig.\ref{Fig4a} it looks that at low temperature the thermal complexity
satisfies the power law
\begin{equation}\label{eqTdV1}
 \delta\bar{\mathcal{V}}^s_d({\bar{T}}) \propto T^\alpha \,.
\end{equation}
This has been checked in Fig.\ref{Fig4b} and $\alpha \simeq 2$. To what extend is this integer exponent robust and what is a physical picture behind it if any? To answer these questions we have first performed more numerical analysis and investigated the dependence of $\alpha$ on the charge $q$.

\begin{figure}[]
 \centering
     {\includegraphics[width=8cm]{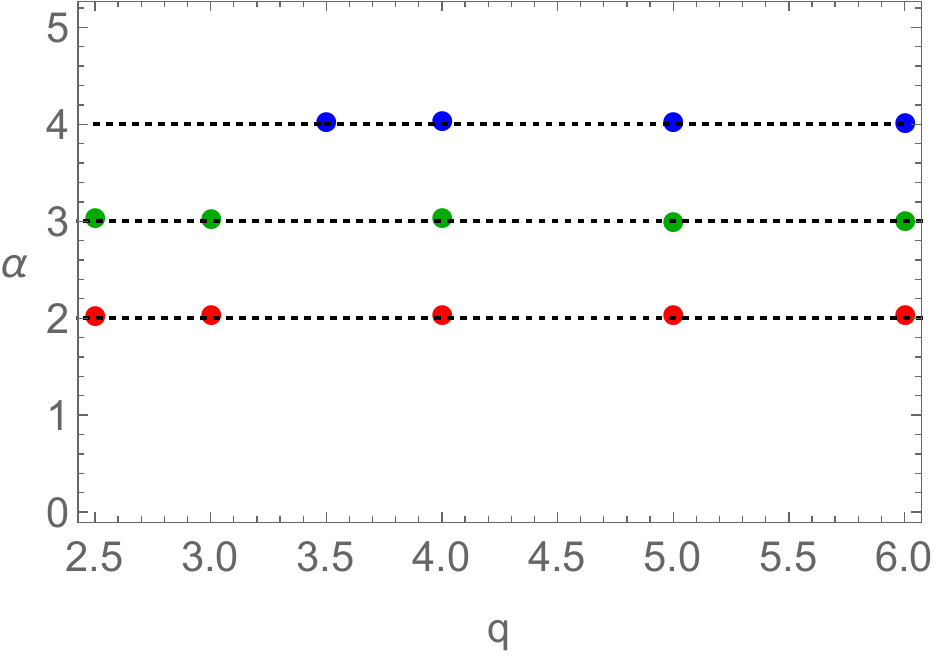}\label{Fig5}}
          \caption{The exponent $\alpha$ as a function of $q$ for  $m^2=0$. Color dots are numerical data. $d=3$(red) $d=4$(green) $d=5$(blue).} \label{Fig5}
\end{figure}

The red dots in Fig.\ref{Fig5} are numerical data of the exponent $\alpha$ which belongs to the range $2\pm2\times10^{-3}$ for the charges $q=2.5,3,3.5,4,4.5,5,6$.  Note that $q^2>3/4$ so that the scalar field can condense spontaneously \cite{Horowitz:2009ij}. In addition, we consider  higher dimensions. In Fig.\ref{Fig5} the green dots are for $d=4$ and the blue dots are for $d=5$ case. Based on our numerical results, we may deduce that the exponent $\alpha$ is independent of the $q$ and
\begin{equation} \label{m20}
\alpha = d-1 \,.
\end{equation}

Indeed, this simple and universal exponent can be understood by the near horizon geometry of the holographic superconductor at low temperature. For $m^2=0$, it is shown to be the AdS-Schwarzschild black hole~\cite{Basu:2011np}:
\begin{equation}\label{Extremal1}
\begin{split}
\chi(z) = \chi_{0}       \,, \quad       f(z) = 1 - \tilde{z}^d      \,, \quad      A_t(z) = 0       \,,\quad     \phi(z) = \phi_{0} \,
\end{split}
\end{equation}
where $\chi_{0}$ and $\phi_{0}$ are constant. It can be also seen from our numerical solutions in Fig.\ref{m0PLOT2b}, where the dashed black curve is the AdS-Schwarzschild blackening factor, $f(\tilde{z}) = 1-\tilde{z}^3$. We see that as the temperature goes down the curves approach $f(\tilde{z}) = 1-\tilde{z}^3$ near horizon.

Therefore, the thermal complexity of the superconductor with $m^2=0$ can be approximately computed as the complexity of formation of the AdS-Schwarzschild black hole:
\begin{equation} \label{under1}
\bar{\mathcal{V}}_{\text{AdS$_{d+1}$-Schwarzschild black hole}}-\bar{\mathcal{V}}_0  \propto T^{d-1} \,,
\end{equation}
which has been computed in \cite{Chapman:2016hwi,Kim:2017qrq}. Note that the geometry \eqref{Extremal1} is valid only near horizon so the volume integral based on \eqref{Extremal1} is not precise for \eqref{deltaVs}. Here, we are assuming that the correction due to this discrepancy are common to both terms in \eqref{deltaVs} and cancelled out.  This is justified a posteriori by \eqref{m20}.

\subsubsection{Complex scalar mass: $m^2\neq0$}

What if we consider $m^2 \neq 0$? In this case, there is no known analytic small temperature geometry for holographic superconductor so far. Thus, for $m^2 \neq 0$ it is not easy to have  analytic understanding such as \eqref{m20} and \eqref{under1}.
However, the {\it zero} temperature solution for holographic superconductor has been analyzed in \cite{Horowitz:2009ij}.  It shows that the ground state geometry depends on $d$, $q$ and $m^2$ in general. Therefore, we may expect that the scaling behavior of the thermal complexity depends on $d$ and $q$ for a given $m^2$. Note that only if $m^2=0$, the zero and small temperature geometry of the holographic superconductor is independent of $q$ as shown in \eqref{Extremal1}. It is why the thermal complexity in that case does not depend on $q$ and only a function of $d$.

Even though we don't have an analytic intuition from the ground state geometry for $m^2 \neq 0$, we can still analyze the scaling behavior of the thermal complexity at low temperature numerically and try to understand physics behind it.

\begin{figure}
  \centering
     \subfigure[$\Delta\bar{\mathcal{V}}_3$ and $\delta\bar{\mathcal{V}}_3$ vs $T/T_c$ ]
  {\includegraphics[width=6.2cm]{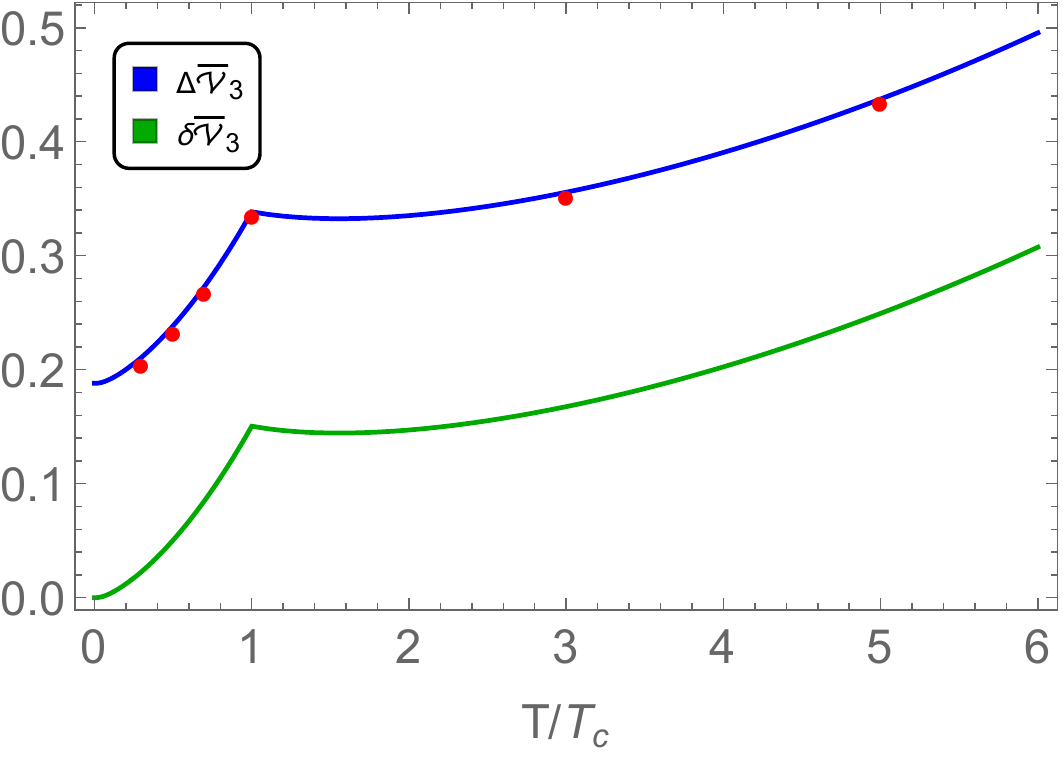}\label{Fig7a}} \ \  \
     \subfigure[$\ln(\delta\bar{\mathcal{V}}_3^\mathrm{s})$ vs. $\ln(T/T_c)$]
  {\includegraphics[width=7.7cm]{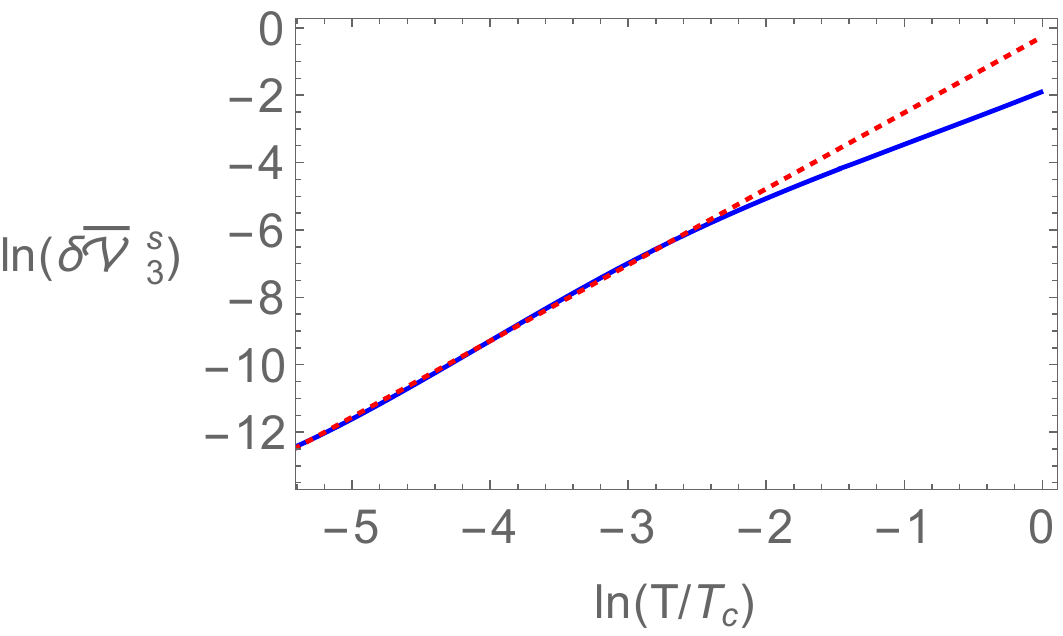}\label{Fig7b}}
  \caption{The complexity of formation and thermal complexity of the holographic superconductor with { $m^2=-2$ and $q=1$. $T_c/\mu\simeq 0.0208 $ is the critical temperature.} } \label{Fig7}
  \end{figure}

\begin{figure}[]
 \centering
     \subfigure[$m^2=-2$. The fitting curve: $\alpha \simeq\, 2.2 + q^2/20$ ]
     {\includegraphics[width=7.1cm]{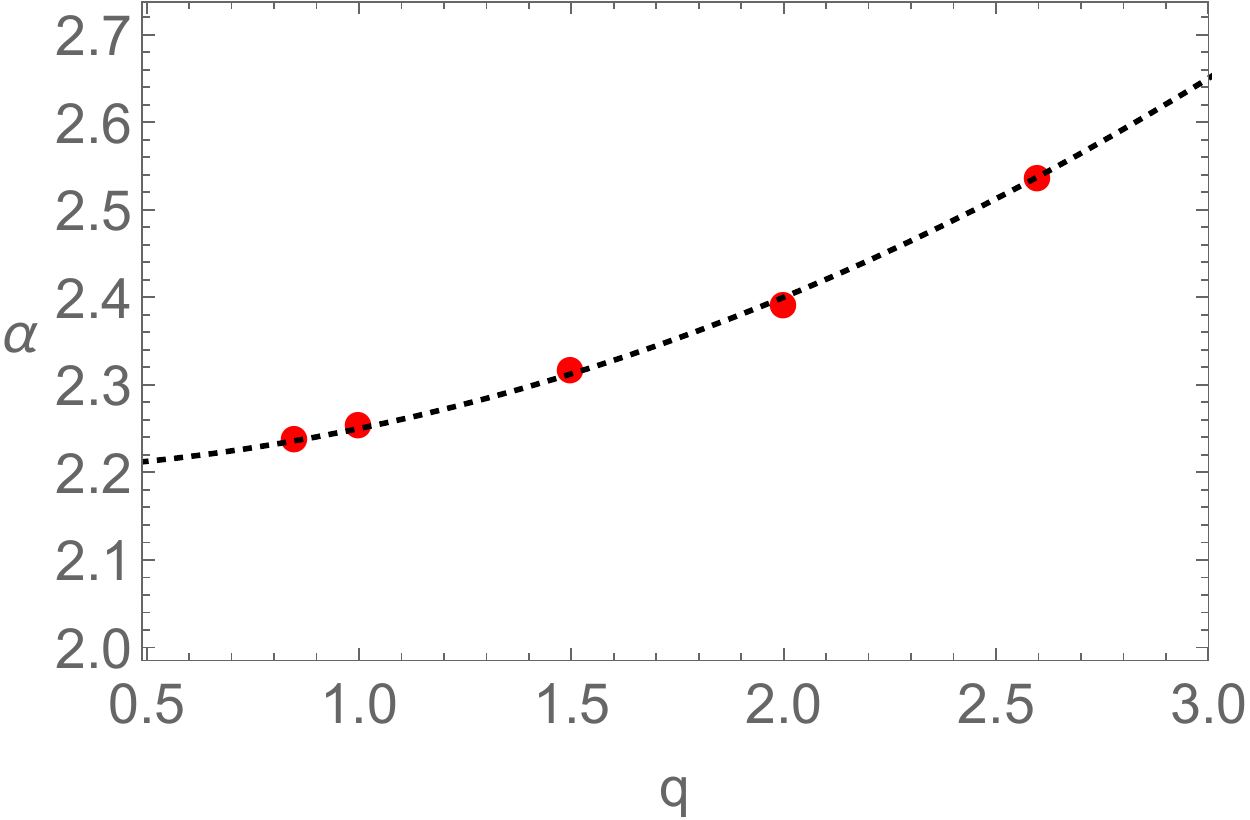} \label{Fig8a}}
     \subfigure[$m^2=1/2$. The fitting curve: $\alpha \simeq\, 2+  \frac{1}{7q^2}$]
     {\includegraphics[width=7.1cm]{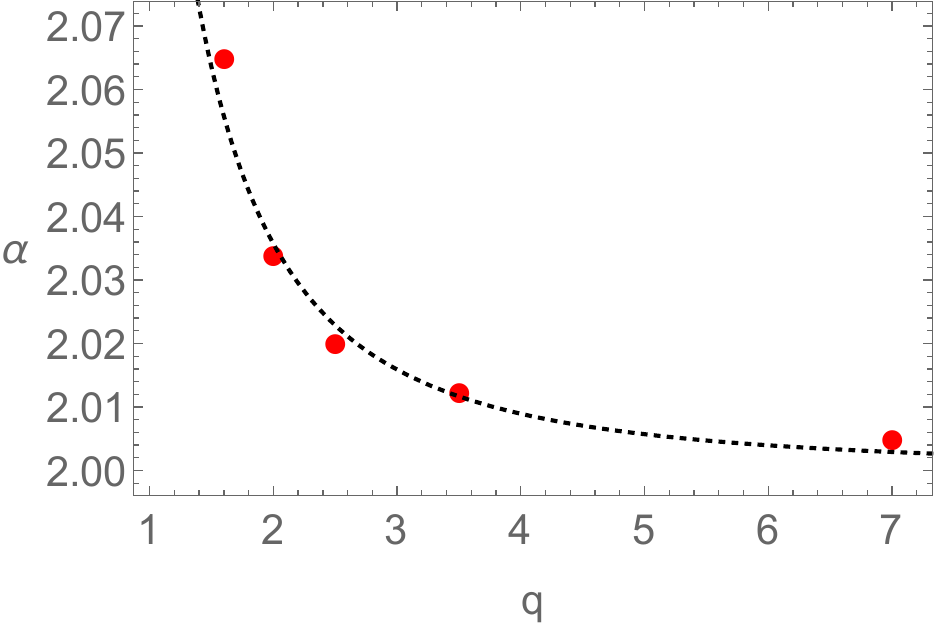} \label{Fig8b}}
          \caption{The exponent $\alpha$ vs $q$ for $m^2\neq0$. Red dots are numerical data. The dashed lines are the fitting curves in \eqref{approxqalpha1a} and \eqref{approxqalpha1b}. } \label{Fig8}
\end{figure}

To be specific, let us consider $d=3$. For a negative mass squared we choose $m^2 =-2$ and for a positive mass squared we choose $m^2 = 1/2 $. First, for $m^2 =-2$ Fig. \ref{Fig7a} shows the complexity of formation and thermal complexity with $q=1$ and from Fig. \ref{Fig7b} we can read off the exponent $\alpha$.
We repeat this computation by changing $q$ and the results are summarized in Fig. \ref{Fig8a}. Next, for  $m^2 =1/2$, by the same procedure, we obtain the exponent $\alpha$, which is shown in Fig. \ref{Fig8b}.
The numerical data in Fig. \ref{Fig8} can be fitted as
\begin{align}
  \alpha &\simeq\, \alpha_{1} + \alpha_{2} q^2 \,, \qquad (m^2 = -2) \,,  \label{approxqalpha1a} \\
   \alpha  &\simeq\, \alpha_{3}+ \alpha_{4} \, q^{-2} \,, \qquad (m^2 = 1/2) \,, \label{approxqalpha1b}
\end{align}
which are represented as a dashed line in Fig. \ref{Fig8a} with $(\alpha_{1}, \, \alpha_{2}) = (2.2, 1/20)$ and $(\alpha_{3}, \alpha_4 )=(1.998, 1/7)$. We have checked that these behaviors are qualitatively true for other values of $m^2$: For a negative mass squared $\alpha$ increases as $q$ increases while for a positive mass squared $\alpha$ decreases as $q$ increases. It seems that $\alpha_1$ and $\alpha_3$ goes to $d-1$ and $\alpha_2$ and $\alpha_4$ goes to zero as $m^2$ goes to zero, reproducing the results in section \ref{m20case}.

Let us conclude this subsection by a remark on the possible low temperature geometry of the holographic superconductor with $m^2 >0$.
When $m^2>0$, it is known that the extremal geometry is Lifshitz geometry:
\begin{equation}\label{}
\begin{split}
e^{\chi(z)} = z^{2(\zeta-1)}       \,, \quad       f(z) = f_{0}      \,, \quad      A_t(z) = A_{0} \, z^{-\zeta}       \,,\quad     \phi(z) = \phi_{0}
\end{split}
\end{equation}
where
\begin{equation}\label{LifshitzGeo}
\begin{split}
f_{0} = \frac{d(d-1)}{(d-1+\zeta)(d-2+\zeta)}         \,, \quad     A_{0} = \sqrt{\frac{2(\zeta-1)}{\zeta}f_{0}}        \,, \quad    \phi_{0} = \sqrt{\frac{(d-1)\zeta}{2q^2}f_{0}} \,,
\end{split}
\end{equation}
and $\zeta$ is the lifshitz dynamical exponent satisfying
\begin{equation}\label{Lifmq}
\begin{split}
m^2 = \frac{2(\zeta-1)}{\zeta}q^2 \,.
\end{split}
\end{equation}
If $d=3$,  \eqref{LifshitzGeo}  is reduced to the case in \cite{Horowitz:2009ij}.

To follow the same logic as $m^2=0$ case in section \ref{m20case}, we may assume that the low temperature solution is obtained by the emblackening factor:
\begin{equation} \label{nosol}
f(z) = f_0 \left(1 - \tilde{z}^{d-1+\zeta}\right) \,.
\end{equation}
The complexity of formation for this Lifshitz black hole is $  \Delta\mathcal{V} \sim T^\frac{d-1}{\zeta}$  which reduces to AdS-Schwarschild case as $\zeta \rightarrow 1$.
Then the exponent $\alpha$ for Lifshitz black hole is
\begin{equation}\label{AlphaLif}
\begin{split}
\alpha = \frac{d-1}{\zeta} = (d-1) - \frac{(d-1)m^2}{2 q^2} \,,
\end{split}
\end{equation}
where \eqref{Lifmq} is used.  For $d=3$
\begin{equation}
\alpha = 2 - \frac{m^2}{q^2} \,,
\end{equation}
which is different from our numerical fitting \eqref{approxqalpha1b}: $\alpha_3$ is similar but the sign of $\alpha_4$ is opposite. This is because the simple emblackening factor \eqref{nosol} is not the low temperature solution of the holographic superconductor for $m^2>0$.


\section{Time-dependent complexity}

\begin{figure}[]
 \centering
     {\includegraphics[width=13cm]{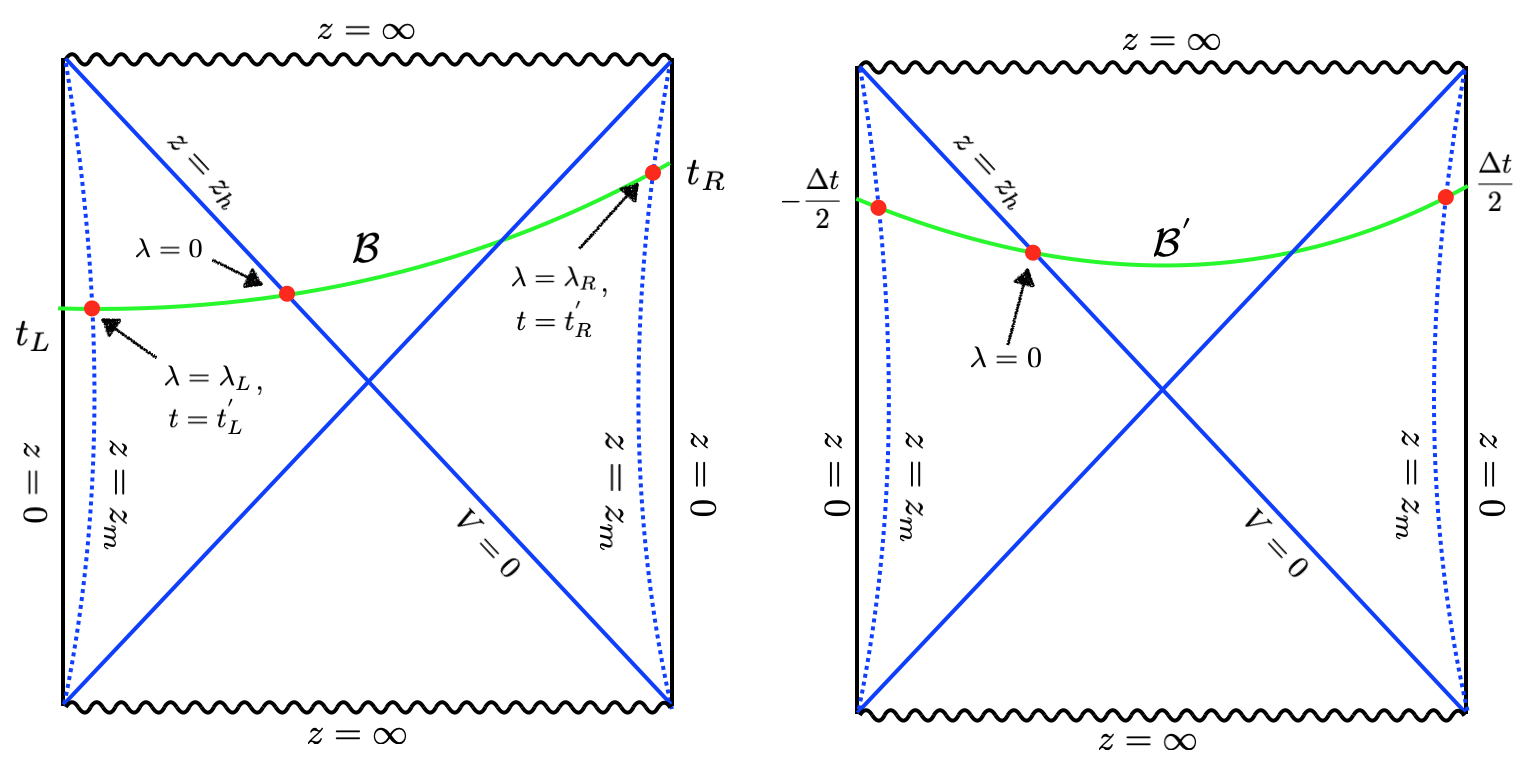} \label{}}
          \caption{  The maximal volume surface in the AdS-Schwarzschild geometry.  At two boundaries, $t_L$ and $t_R$ stand for two states dual to the states in TFD. $z_h$ is the horizon radius. $\lambda$ is used to parametrize $z(\lambda)$ and $V(\lambda)$. $\mathcal{B}$ and $\mathcal{B'}$ are schematic representation of the maximum codimension-one bulk surfaces connecting $t_L$ and $t_R$ in the left plot, $-\frac{\Delta t}{2}$ and $\frac{\Delta t}{2}$ in the right plot, respectively.  } \label{FigAdS}
\end{figure}

In this section, we study the time-dependent complexity or complexity of formation.
We consider the same ansatz as \eqref{anstz1}:
\begin{equation}\label{metricAdS}
\td s^2 = {1 \over z^2} \left[ -f(z)e^{-\chi(z)}\td t^2+f^{-1}(z)\td z^2+     \sum_{i=1}^{d-1}\td x^2_i    \right].
\end{equation}
From the CV conjecture perspective, it means that
we evaluate the volume of the maximum codimension-one bulk surface ($\mathcal{B}$ or $\mathcal{B'}$) in Fig.\ref{FigAdS} instead of the horizontal surface in Fig.\ref{Fig1} or Fig.\ref{FigP2}. \footnote{For the purpose of displaying the maximal surface at  finite time we use the Penrose diagram for the AdS Schwarzschild geometry in Fig.\ref{FigAdS}.} The green curves $\mathcal{B}$ or $\mathcal{B'}$ are schematic ones. As the metric \eqref{metricAdS} has the time translation symmetry the volume of $\mathcal{B}$ and $\mathcal{B'}$ are the same, if
\begin{equation} \label{Dtrl}
\Delta t \equiv | t_{R} - t_{L} |
\end{equation}

\subsection{Method of computation} \label{genmethod}

To compute the volume of the codimension-one bulk surface let us first introduce the null coordinate $V(t,z)$:
\begin{align}\label{Nullcoordi}
\begin{split}
	  V(t,z)  &=  e^{\beta \, v(t,z)}  =  e^{\beta \, (t - z^*(z))} \,, \\
	  z^*(z)  &=  \int_{0}^{z} \frac{e^{\chi(\tilde{z})/2}}{f(\tilde{z})} \td \tilde{z}
	   \,,
\end{split}
\end{align}
where $v(t,z)$ is the infalling Eddington-Finkelstein coordinate and $\beta$ is determined by the near horizon expansion of the equations of motion.
In this coordinate system, the metric \eqref{metricAdS} becomes
\begin{align}\label{Nullmetric}
\begin{split}
	\td s^{2} &=  \frac{1}{z^2} \left[ - f(z) e^{-\chi(z)} \td v^{2}  - 2 e^{-\chi(z)/2} \td v \td z +  \sum_{i=1}^{d-1}\td x^2_i \right] \,,\\
	              &=  \frac{1}{z^2} \left[ - \frac{f(z) e^{-\chi(z)}}{\beta^2V^2} \td V^{2} - \frac{2 e^{-\chi(z)/2}}{\beta V}\td V \td z + \sum_{i=1}^{d-1}\td x^2_i \right] \,.
\end{split}
\end{align}

By considering the induced metric on the green curves parameterized by $\lambda$ in Fig. \ref{FigAdS}.
\begin{align}\label{Nullmetric}
\begin{split}
	\td s_{c}^{2} &:=  \frac{1}{z(\lambda)^2} \left[ \left(- \frac{f(z(\lambda)) e^{-\chi(z(\lambda))}}{\beta^2V(\lambda)^2}V'(\lambda)^{2}  -  \frac{2 e^{-\chi(z(\lambda))/2}}{\beta V(\lambda)}V'(\lambda)z'(\lambda) \right) \td \lambda^2  +  \sum_{i=1}^{d-1}\td x^2_i \right] \,.
\end{split}
\end{align}
we have the volume of the codimension-one bulk surface
\begin{align}
	\mathcal{V}  &=  \Omega_{d-1} \int \frac{1}{z(\lambda)^{d}} \sqrt{  - \frac{f(z(\lambda)) e^{-\chi(z(\lambda))} }{\beta^2V(\lambda)^2} V'(\lambda)^{2}  -  \frac{2 e^{-\chi(z(\lambda))/2} }{\beta V(\lambda)} V'(\lambda)z'(\lambda)  } \td \lambda \label{Lag1Vol}  \\
	& \equiv  \Omega_{d-1} \int \mathcal{L}_1\, \td \lambda\,, \label{Lag1Vol1}
\end{align}
where $\Omega_{d-1} \equiv \int \td \Omega^{d-1}$ is the volume of the spatial geometry.
We obtain only one Euler-Lagrangian equation from \eqref{Lag1Vol}
\begin{align}\label{Lag1eom}
\begin{split}
	& -  \frac{z''}{z'} + \frac{(4d + z\chi'(z))z'}{2z} + \frac{V''}{V'} + \frac{ e^{-\chi(z)} f(z) (-z f'(z) + f(z)(2d + z \chi'(z)) ) }{2 \beta^2 V^2 z z'}V'^{2} \\
	& + \frac{ -2 \beta z + e^{-\chi(z)/2}(-3 z f'(z) + 3 f(z)(2d + z \chi'(z))) }{2 \beta V z}V' = 0 \,.
\end{split}
\end{align}

However, we need two equations for two independent fields($z(\lambda), V(\lambda)$). To resolve this issue, we introduce the auxiliary field $\varepsilon(\lambda)$  to the volume integral.
\begin{align}\label{Lag2eom}
	\mathcal{V}  =  \Omega_{d-1} \int \left( \frac{1}{\varepsilon(\lambda)}\mathcal{L}_1^2 + \varepsilon(\lambda) \right) \td \lambda  \equiv  \Omega_{d-1} \int \mathcal{L}_2\, \td \lambda\,,
\end{align}
from which, we have three Euler-Lagrangian equations:
\begin{align}\label{EL2}
\begin{split}
	& \frac{e^{\chi(z)/2}\beta}{f(z)}z'' - \frac{e^{\chi(z)/2}\beta(4d + z \chi'(z))}{2 f(z) z}z'^{2} - \frac{e^{\chi(z)/2}\beta \varepsilon'}{\varepsilon f(z)} z' + \frac{V''}{V} - \frac{V'^{2}}{V^2} \\
	& \qquad \qquad + \left( -\frac{\varepsilon'}{\varepsilon V} + \frac{z z' f'(z) - z' f(z)(2d + z\chi'(z)) }{f(z) V z} \right)V' = 0 \,,  \\
	& V'' - \frac{2\beta z + e^{-\chi/2}(z f'(z) - f(z)(2d + z \chi'(z)))}{2\beta V z} V'^{2} - \frac{\varepsilon' V'}{\varepsilon} = 0 \,, \\
	& 1 + \frac{z^{-2d}}{\varepsilon^2}\left( \frac{2 e^{-\chi/2} V' z'}{\beta V} + \frac{e^{-\chi} f(z) V'^2}{\beta^2 V^2} \right) = 1- \frac{\mathcal{L}_1^2}{\varepsilon^2} =0 \,.
\end{split}
\end{align}
Among these, only two are independent and we will choose the second and third equations as independent ones.
From here, we take $\varepsilon(\lambda)$=1, recovering the original variational problem. The two independent equations read
\begin{align}
	& V'' - \frac{2\beta z + e^{-\chi/2}(z f'(z) - f(z)(2d + z \chi'(z)))}{2\beta V z} V'^{2} = 0 \,, \label{eqbtz1} \\
	& 1 + z^{-2d}\left( \frac{2 e^{-\chi/2} V' z'}{\beta V} + \frac{e^{-\chi} f(z) V'^2}{\beta^2 V^2} \right) = 1- \mathcal{L}_1^2 =0 \,. \label{eqbtz2}
\end{align}
We will solve the equation of motions \eqref{eqbtz1} and \eqref{eqbtz2}  numerically starting from the horizon at $(V(0), z(0)) = (0, z_h)$. See the red dot for $\lambda=0$ in Fig.\ref{FigAdS}. Near this point the series solutions are obtained as
\begin{align}\label{bcbtz}
\begin{split}
	& V = V^{(1)}\lambda - \frac{f'(z_{h})(4d + 3 z_{h}\chi'(z_{h})) - 2 z_{h} f''(z_{h} ) }{4 z_{h} f'(z_{h})} z^{(1)} V^{(1)} \lambda^2 + \cdots \,,\\
	& z = z_h + z^{(1)} \lambda + \frac{z_{h}^{1+2d}f'(z_{h}) + (4d + z_{h}\chi'(z_{h})) {z^{(1)}}^2 }{4 z_{h}} \lambda^2 + \cdots \,,
\end{split}
\end{align}
where $\beta$ introduced in \eqref{Nullcoordi} is determined as
\begin{equation}
\beta =  -\frac{1}{2}e^{-\chi(z_{h})/2} f'(z_{h}) \,,
\end{equation}
which is nothing but $2\pi T$  as expected in \eqref{HawkingT}.

From here we introduce again $\tilde{z} = z/z_h$ defined in \eqref{ztilde} without loss of generality.
For a given initial values ($V^{(1)}$, $z^{(1)}$) the solutions ($\tilde{z}(\lambda), V(\lambda)$) are numerically determined. Thus, at the cut-off $\tilde{z}_m$ (see Fig.\ref{FigAdS}),  ($\lambda_{R}$, $\lambda_{L}$) can be obtained as
\begin{align}\label{lambdaval}
\begin{split}
   \tilde{z}(\lambda_{R}) = \tilde{z}_{m}   \,, \qquad    \tilde{z}(\lambda_{L}) = \tilde{z}_{m}   \,.
\end{split}
\end{align}
Note that $\lambda_R>0$ and $\lambda_L<0$.
Once we know the parameters ($\lambda_{R}$, $\lambda_{L}$) at the cut-off $\tilde{z}_m$, the complexity \eqref{Lag1Vol1} is computed straightforwardly as follows.
\begin{align}\label{time111}
\begin{split}
\mathcal{V}  = \Omega_{d-1} \int \mathcal{L}_1\, \td \lambda  =  \Omega_{d-1} \int \, \td \lambda = \Omega_{d-1}(\lambda_{R} - \lambda_{L}) \,.
\end{split}
\end{align}
where, the second equality holds because $\mathcal{L}_1=1$, following from \eqref{eqbtz2}.

The times at the boundary cut-off $\tilde{z}_m$ \eqref{lambdaval} are obtained from \eqref{Nullcoordi}:
\begin{align}\label{tLRval}
\begin{split}
t_{R}^{'}  \equiv t(\lambda_{R}) =  \frac{\log(V(\lambda_{R}))}{\beta} + z^{*}(\lambda_{R})  \,, \qquad  t_{L}^{'}   \equiv t(\lambda_{L})  =  \frac{\log(V(\lambda_{L}))}{\beta} + z^{*}(\lambda_{L}) \,.
\end{split}
\end{align}
Thus, from \eqref{Dtrl}
\begin{equation}
\Delta t = | t_{R}^{'} - t_{L}^{'} | =  \frac{1}{\beta}\left(\log(V(\lambda_{R})) - \log(V(\lambda_{L}))\right) \,,
\end{equation}
where $z^{*}(\lambda_{R}) - z^{*}(\lambda_{L}) = 0$ by \eqref{Nullcoordi} and \eqref{lambdaval}. Strictly speaking, $\Delta t$ is the value in the limit of $\tilde{z}_m \rightarrow 0$, but for our numerics, $\Delta t$ is defined at $\tilde{z}_m = 10^{-2}$.

In summary, from \eqref{time111} and \eqref{tLRval}, the time dependent complexity of formation \eqref{barFormula1} or \eqref{super11} reads:
\begin{align}
     \Delta\bar{\mathcal{V}}_d(\bar{T}, \Delta t)  &=  \frac{\Delta\mathcal{V}}{\mu^{d-1}\Omega_{d-1}} = \frac{1}{\tilde{\mu}^{d-1}} \left[  \lambda_{R}(\tilde{z}_{m}) - \lambda_{L}(\tilde{z}_{m}) - \frac{2}{(d-1)}\frac{1}{\tilde{z}^{d-1}_{m}}  \right] \,,  \label{xxx1}\\
     \Delta t &=  \frac{1}{\beta}\left[\log\frac{V(\lambda_{R}(\tilde{z}_m))}{V(\lambda_{L}(\tilde{z}_m))}\right] \,,
\end{align}
where the last term in \eqref{xxx1} comes from $\mathcal{V}_0$ in \eqref{CVAdS1}.  Note that we do not use the superscript `n' or `s' here compared with \eqref{barFormula1} and \eqref{super11} because \eqref{xxx1} is valid for both normal and superconducting phase.

%

%

\subsection{Results and Lloyd's bound}
For numerical analysis we consider the (3+1) dimensional holographic superconductor model based on \eqref{SetupModel} with $q=1$ and $m^2=-2$. We first construct the superconductor solutions: $f(\tilde{z}), \chi(\tilde{z}), A_t(\tilde{z}), \phi(\tilde{z})$, similar to Fig. \ref{m0PLOT2}.

\begin{figure}[]
 \centering
 \subfigure[$\tilde{z}(\lambda)$. $\lambda_{R}=5002.31$ and $\lambda_{L}=-5001.73$ at the cut-off $\tilde{z}_m = 10^{-2}$.  ]
     {\includegraphics[width=7.3cm]{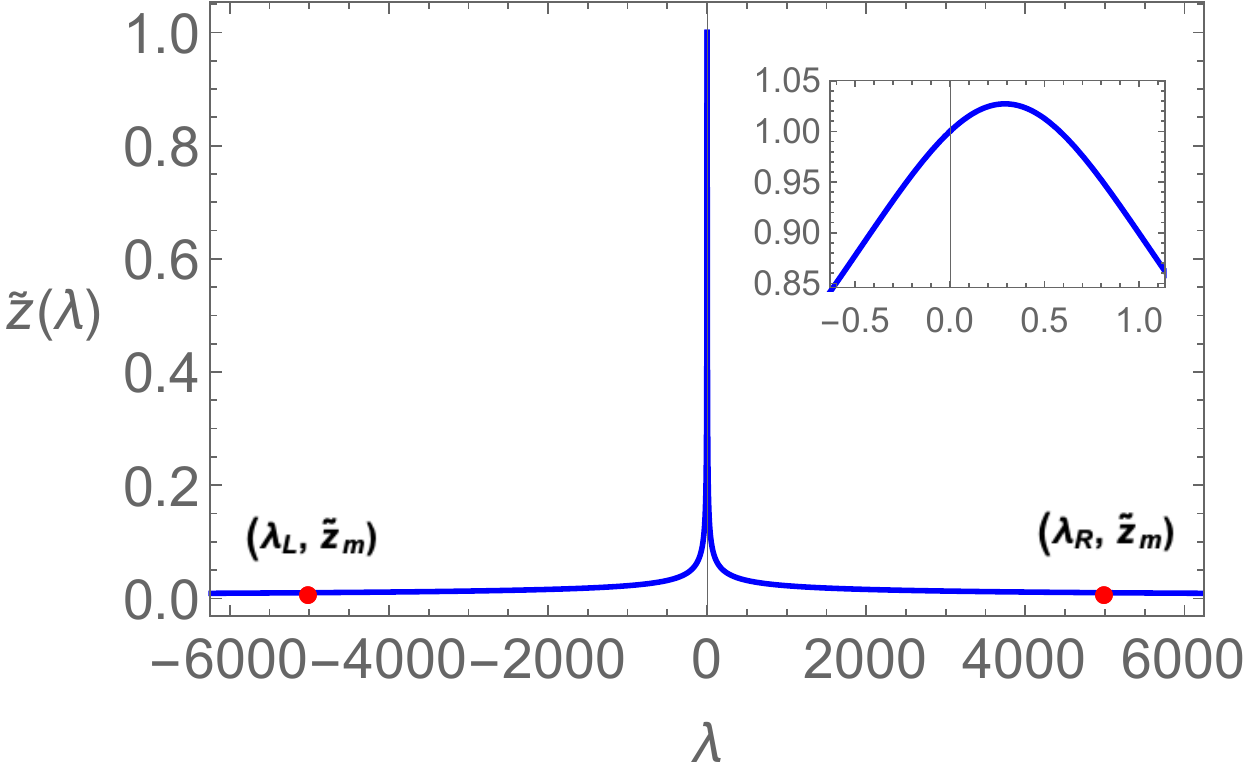}} \ \ \
\subfigure[$V(\lambda)$. $V(\lambda_{R})= 3.247$ and $V(\lambda_{L})=-0.688 $ at the cut-off $\tilde{z}_m = 10^{-2}$, which are out of range of the plot. ]
     {\includegraphics[width=7.1cm]{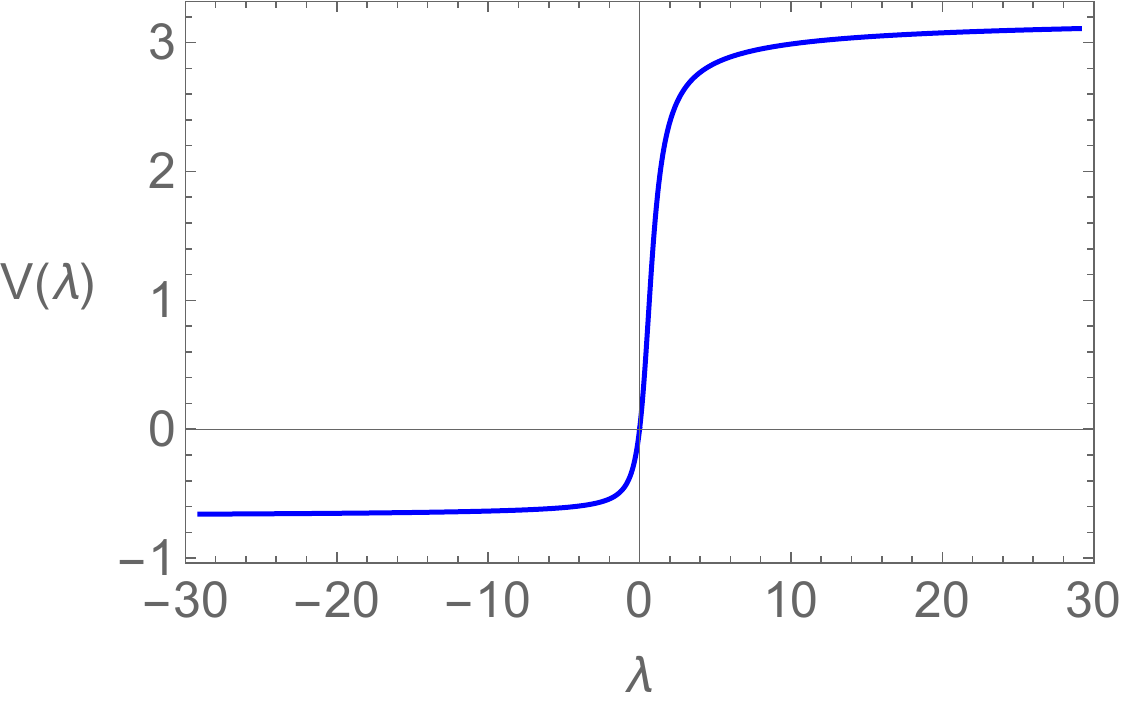}}
          \caption{The numerical solutions: $\tilde{z}(\lambda), \, V(\lambda)$ for $T/T_{c}=0.7$. This plot is generated by the initial data $(z^{(1)}, \, V^{(1)}$ = (0.175, 1).
} \label{Solset3}
\end{figure}

\begin{figure}[]
 \centering
     {\includegraphics[width=8cm]{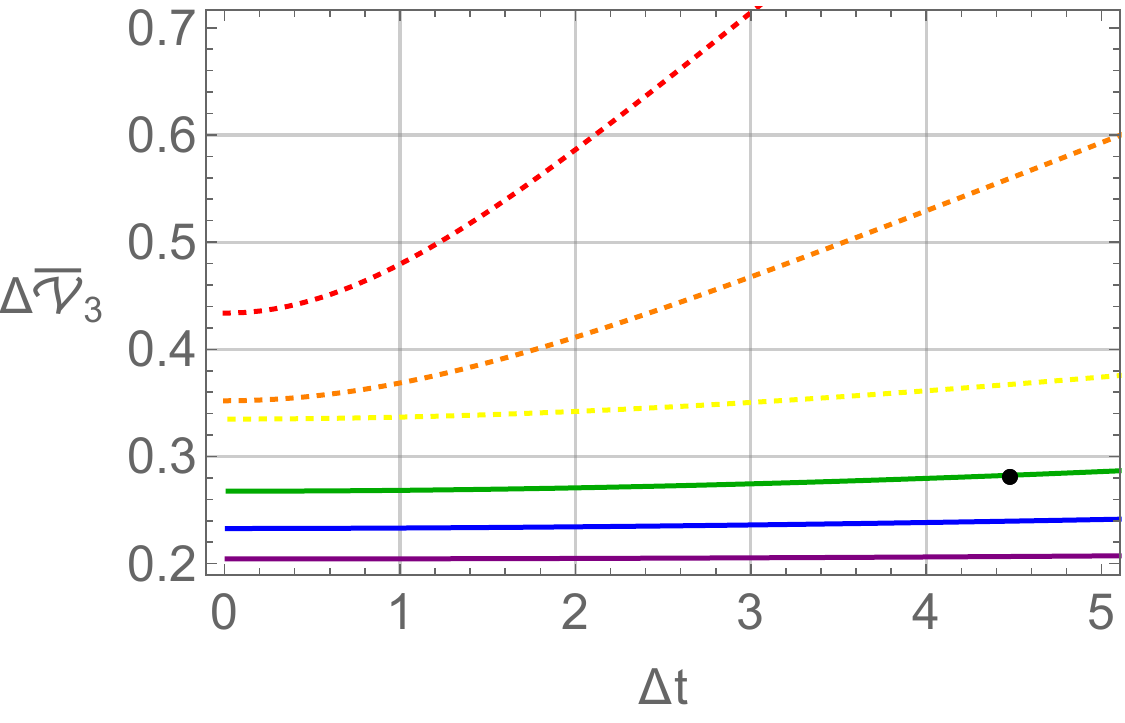} }
          \caption{The time-dependent complexity of formation of the holographic superconductor model. Various colors represent different temperatures. i.e. $T/T_{c} = 5, 3, 1, 0.7, 0.5, 0.3$ (red, orange, yellow, green, blue, purple). Dotted curves are for normal phase and solid curves are for superconducting phase.} \label{TimeDepComResult11}
\end{figure}

As explained in the previous section, for a given set of initial values $(z^{(1)}, V^{(1)}) = (0.175, 1)$, the functions $z(\lambda)$ and $V(\lambda)$ are obtained by numerically solving \eqref{eqbtz1} and \eqref{eqbtz2} with initial conditions \eqref{bcbtz}. For the superconducting phase at $T/T_c =0.7$ the result is displayed in Fig. \ref{Solset3}.
At the boundary cut-off $\tilde{z}_m=10^{-2}$ we obtain $\lambda_{R}=5002.31$ and $\lambda_{L} = -5001.73$ by \eqref{lambdaval}. Furthermore, $V(\lambda_R) = 3.247$ and $V(\lambda_L) = -0.688$ so $t_{R}=3.40$ and $t_{L}=-1.07$ by \eqref{tLRval} with $\beta=0.346$. As a result, the complexity of formation is $ \Delta\bar{\mathcal{V}}_d(\bar{T}, \Delta t) = 0.282$, where $\tilde{\mu}$ is computed as $A_t (\tilde{z}_m) = 3.788$.  It is marked as a black dot in Fig.\ref{TimeDepComResult11}. By changing the initial data set we can complete a green curves in  Fig.\ref{TimeDepComResult11}.

By repeating this procedure for different temperatures, $T/T_c = 5,3,1,0.5,0.3$ we obtain the red, orange, yellow, blue, and purple curves respectively. Note that our numerical method works also for the normal phase ($T>T_c$). The complexity of formation at $\Delta t =0$ agrees with the colored dots in Fig. \ref{Fig7a}, which serves as a consistency check of our numerical analysis.

\begin{figure}[]
 \centering
 \subfigure[Normal phase]
     {\includegraphics[width=7.1cm]{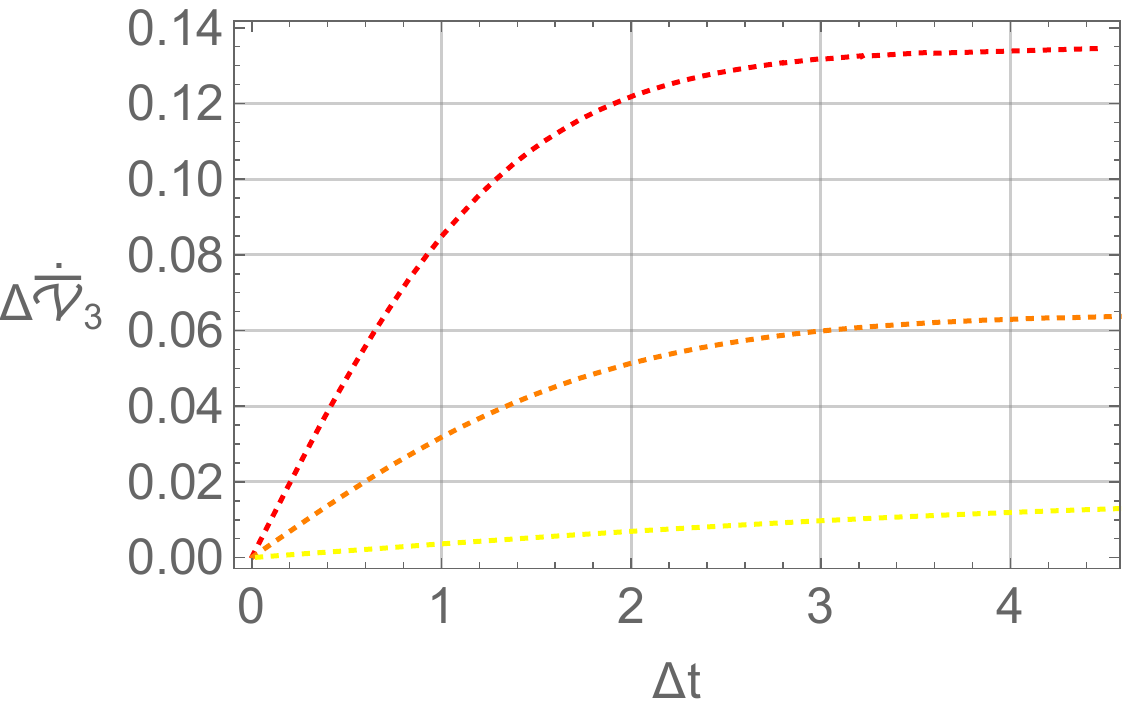} \label{}} \ \ \
\subfigure[Superconducting phase]
     {\includegraphics[width=7.25cm]{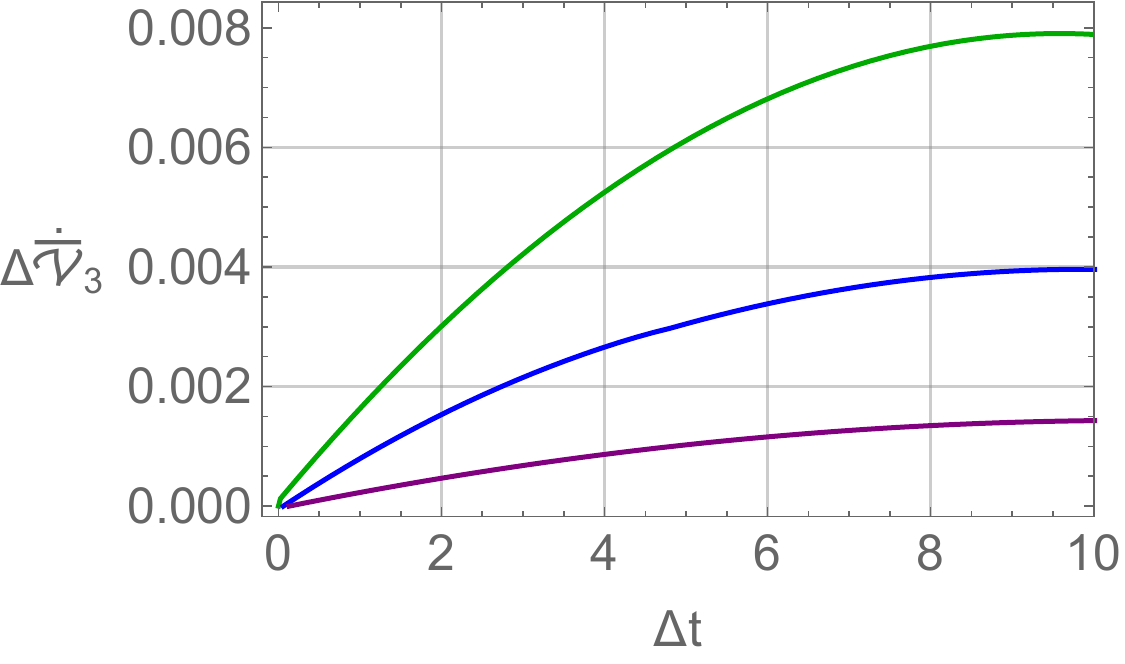} \label{}}
          \caption{The growth rate of the complexity for the holographic superconductor model. Various colors represent different temperatures. i.e. $T/T_{c} = 5, 3, 1, 0.7, 0.5, 0.3$ (red, orange, yellow, green, blue, purple).} \label{TimeDerivativeResult1}
\end{figure}

To see the growth rate of the complexity\footnote{The growth rate of the complexity and the growth rate of the complexity of formation is the same because $\mathcal{V}_0$ is constant in time.},
\begin{equation}
\Delta \dot{\bar{\mathcal{V}}}_d := \frac{\partial \Delta {\bar{\mathcal{V}}}_d}{\partial \Delta t} \,,
\end{equation}
we make the plots for the time derivatives of the complexity in Fig. \ref{TimeDerivativeResult1}.
The higher the temperature is, the bigger the growth rate is. Thus, for a clear presentation we make two plots: (a) normal phase and (b) superconducting phase. We find that there are the upper bounds on the time derivative of complexity for both the normal phase and superconducting phase.
\begin{figure}[]
 \centering
 \subfigure[Normal and superconducting phase]
     {\includegraphics[width=7cm]{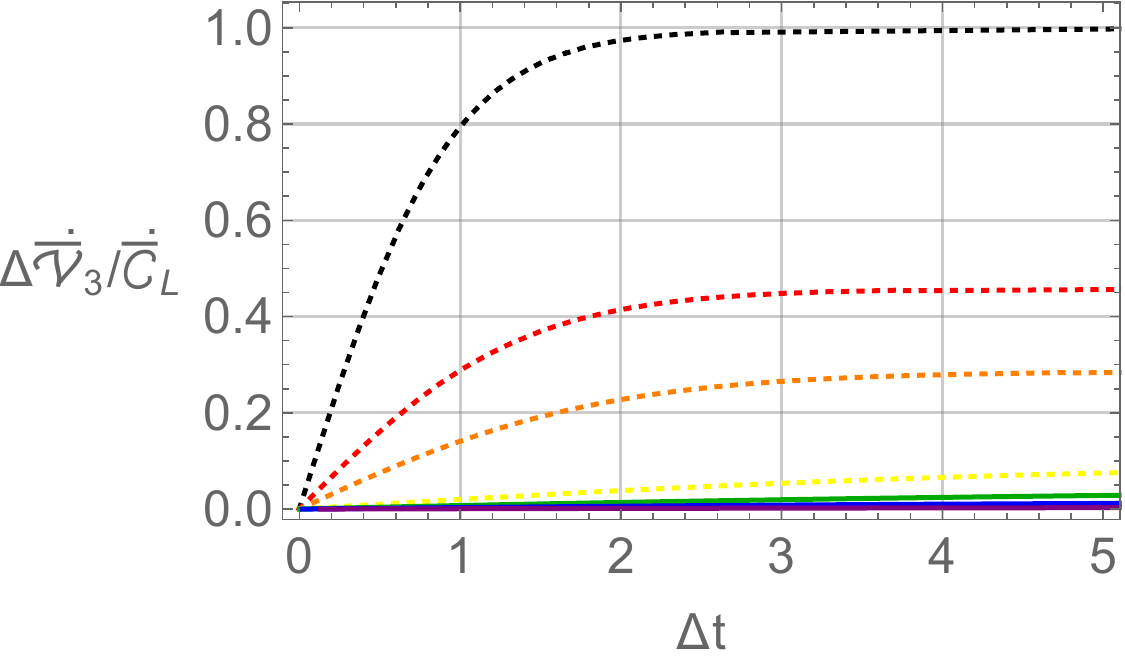} \label{TimeDerivativeResult2a}} \ \
 \subfigure[The closeup of (a) for superconducting phase]
     {\includegraphics[width=7.4cm]{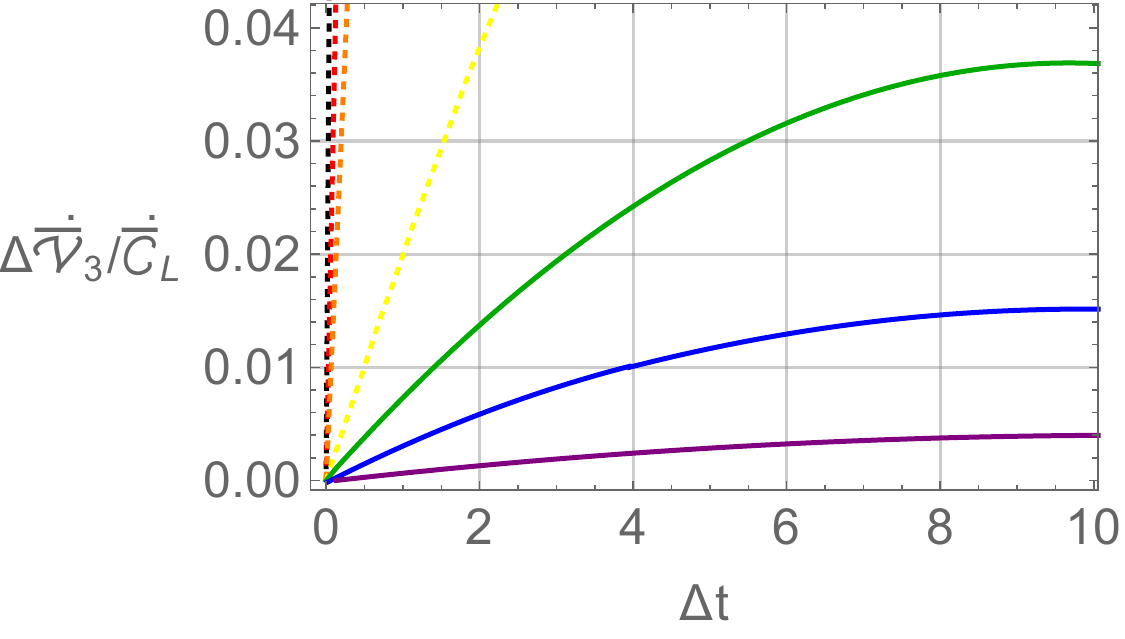} \label{TimeDerivativeResult2b}}
          \caption{The comparison between the growth rate of the complexity for the holographic superconductor model and the Lloyd's bound. Various colors represent different temperatures. i.e. $T/T_{c} = 100, 5, 3, 1, 0.7, 0.5, 0.3$ (black, red, orange, yellow, green, blue, purple). The black curve at high temperature is approaching to the case of AdS-Schwarzschild as shown in \cite{Kim:2017qrq}.
} \label{TimeDerivativeResult2}
\end{figure}

{To compare the upper bound with the Lloyd's bound, we first need to define the  Lloyd's bound, which is proportional to the `mass' (or energy) of the black hole.} However, it is subtle how to define mass in curved spacetimes. For an asymptotic AdS spacetime, there are several different ways to define mass, such as the ADM mass ($M_{\text{ADM}}$), the Komar mass  ($M_{\mathrm{\mathrm{Komar}}}$) and the mass ($M_{\text{h}}$)  based on the holographically renormalized stress tensor. These three definitions give the same results for normal phase but different results for superconducting phase.
See appendix~\ref{Defmass} for details.
{In many cases, $M_{\text{h}}$ is used to define the mass of a boundary state. It has an advantage that the counterterms are fixed according to the ``holographic renormalization''~\cite{Skenderis:2002wp,Papadimitriou:2016yit}. However, $M_{\text{h}}$  may not be appropriate in the context of the Lloyd's bound because it can be negative in general. In the context of the Lloyd's bound, the other two masses have their own advantages.} In the proofs of the ``positive energy theorem''~\cite{Gibbons1983}, the mass of an asymptotic AdS black hole is $M_{\text{ADM}}$. In Ref.~\cite{Yang:2016awy}, it was shown that the mass corresponding to the Lloyd's bound in the late time limit for the CA conjecture is $M_{\mathrm{\mathrm{Komar}}}$.

It is not clear which mass corresponds to the mass in the Lloyd's bound. We check the Lloyd's bound by all  three definitions of mass. In our model it turns out that $M_{\text{ADM}}=M_{\text{h}}\leq M_{\mathrm{\mathrm{Komar}}}$ (See appendix~\ref{Defmass}). Thus,  if the Lloyd's bound is satisfied by the ADM mass, it will be satisfied by any of  three masses. For this reason we focus on the ADM mass.

The Lloyd's bound based on ADM mass is given by
\begin{align}\label{KomarLloyd}
\begin{split}
\dot{\bar{\mathcal{C}}}_{\mathrm{L}}  &:=  \frac{2 \dot{\mathcal{C}}_{\mathrm{Lloyd}}}{\mu^{d-1}\Omega_{d-1}}\,, \qquad  \dot{\mathcal{C}}_{\mathrm{Lloyd}}  := \frac{8\pi  M_{\mathrm{ADM}}}{d-1} \,,
\\
M_{\mathrm{ADM}}  &= -\frac{(d-1) \Omega_{d-1}}{16 \pi}\lim_{z\rightarrow0}(f-1)z^{-d}\,.
\end{split}
\end{align}
%
%
{There is an ambiguity in the definition of the Lloyd's bound. Essentially, it only says $\dot{\mathcal{C}}_{\mathrm{Lloyd}} \sim M_{\mathrm{ADM}}$ and the proportionality constant remains to be fixed.  We choose it as $8\pi/(d-1)$, which is the maximum value for the Schwarzschild blackhole case~\cite{Kim:2017qrq}.  The factors  in the definition of $\dot{\bar{\mathcal{C}}}_{\mathrm{L}}$ is introduced for the same normalization as \eqref{barFormula1} and the factor `2' is necessary to reflect  the contribution from two sides(left and right) of the blackhole. }

The ratio of the growth rate of the complexity to the Lloyd's bound($\dot{\mathcal{C}}_{\mathrm{Lloyd}}$) is presented in Fig. \ref{TimeDerivativeResult2}: (a) is for normal phase and (b) is for superconducting phase. We find that the growth ratio of the complexity becomes smaller than the Lloyd's bound as temperature goes down. This tendency has been observed in normal phase in Fig. 13 of \cite{Carmi:2017jqz}, where the complexity for the normal phase of AdS-RN in (4+1) dimension was studied. Here, we find that it continues to be true also for  superconducting phase.
At a very high temperature (the black curve in Fig. \ref{TimeDerivativeResult2a}) the growth rate of AdS-RN case saturates to the Lloyd's bound, which is similar to the case of the AdS-Schwarzschild as shown in \cite{Kim:2017qrq}. It can be understood by the fact that the high temperature($T/\mu \gg 1$) corresponds to a small chemical potential close to the AdS-Schwarzschild case.

{The Lloyd's bound in the context of the holography was first introduced in the study of the CA conjecture. The complexity by the CA conjecture seemed to saturate to the Lloyd's bound at late time.  It was taken as the first nontrivial evidence supporting the CA conjecture. However, more recent studies~\cite{Moosa:2017yiz,Carmi:2017jqz,Kim:2017qrq,HosseiniMansoori:2018gdu,Ageev:2019fxn} show that such a bound can be violated in many cases.  Only in the late time limit the CA conjecture saturates to the Lloyd's bound {\it from above}. It was first reported by \cite{Carmi:2017jqz,Kim:2017qrq} independently.  It has also been shown that the Lloyd's bound is not satisfied even in AdS-Schwarzschild black holes when time $t_L$ and $t_R$ are small~\cite{Carmi:2017jqz,Kim:2017qrq}.

Our results show that from the perspective of the Lloyd's bound the CV conjecture is more suitable than the CA conjecture as a definition of the holographic complexity.  For AdS-Schwarzschild black holes, RN-AdS black holes and the black holes with nonzero complex scalar hair, we found that the complexity growth rate always satisfies the Lloyd's bound during the whole evolution. Based on these observation we may propose a conjecture: {\it under a few suitable conditions on matter fields or causal structures of spacetime, the complexity by the CV conjecture will always  satisfy the Lloyd's bound.}
}


\section{Conclusions}

In this paper, we have investigated the holographic complexity for the holographic superconductor model in asymptotically AdS$_{d+1}$ spacetime by using the CV conjecture.

We first study a time-independent complexity of formation $\Delta\bar{\mathcal{V}}_d(\bar{T})$ as a function of temperature ($T$) at fixed chemical potential ($\mu$), i.e. $\bar{T} = T/\mu$. A typical behavior of $\Delta\bar{\mathcal{V}}_d(\bar{T})$ is shown in Fig. \ref{Fig4a}.
{In many cases with different masses $m^2$ and charges $q$, we find    that a universal property:  the superconducting phase always has the smaller complexity than the unstable normal phase below the critical temperature.  Thus, it seems that the complexity may play a role of the free energy, indicating the phase transition. It will be  interesting to investigate if this property (higher complexity for unstable phase) is universal for other cases, and understand the reason behind it.}

In superconducting phase, the complexity of formation always decreases as temperature goes down\footnote{In normal phase, the complexity of formation may increase right before the critical temperature if $m^2 < 0$ as shown in Fig. \ref{Fig7a}. This can be understood from Fig. \ref{FIGnormal}.}.
In the high temperature limit, the complexity of formation scales as $\Delta\bar{\mathcal{V}}_d(\bar{T}) \sim T^{d-1}$, which is consistent with the AdS-Schwarzschild Black brane geometry. In the low temperature limit, if the system were in normal phase, the complexity of formation would behave as $\Delta\bar{\mathcal{V}}_d(\bar{T}) \sim \log T$ as shown in Fig. \ref{FIGnormal}. However, at low temperature, the system must be in the superconducting phase and in this case it again scales as $\delta\bar{\mathcal{V}}_d(\bar{T}) \sim T^\alpha$, where $\alpha$ is a function of the dimension $d$, the complex scalar mass squared $m^2$, and the $U(1)$ charge $q$ \footnote{$\Delta\bar{\mathcal{V}}_d(\bar{T}) =  \delta\bar{\mathcal{V}}_d(\bar{T})$ + constant. The constant is just an overall shift so it does not matter in discussing the scaling behavior in temperature. }.

In particular, if $m^2=0$, $\alpha = d-1$ {\it independent} of $q$, which is the same as the high temperature limit. i.e.
\begin{equation}
\delta\bar{\mathcal{V}}_d(\bar{T}) \sim T^{d-1} \,.
\end{equation}
It can be understood by the fact that the near horizon geometry of the holographic superconductor with $m^2=0$ at low temperature is nothing but the AdS-Schwarzschild Black brane geometry.
For $m^2 \ne 0$, our numerical analysis suggests the following behavior (see for example Fig. \ref{Fig8}.).
\begin{align}
  \alpha &\simeq\, \alpha_{1} + \alpha_{2} q^2 \,, \qquad (m^2 < 0) \,,  \label{y11}  \\
   \alpha  &\simeq\, \alpha_{3}+ \alpha_{4} \, q^{-2} \,, \quad \ (m^2 > 0)  \label{y22} \,,
\end{align}
where $\alpha_1$ and $\alpha_3$ goes to $d-1$ and $\alpha_2$ and $\alpha_4$ goes to zero as $m^2$ goes to zero, reproducing the results for $m^2=0$. For $m^2 \neq 0$ it is not easy to have a geometric understanding for $\alpha$ because there is no available near horizon  analytic geometry at low temperature. We leave a better physical understanding of \eqref{y11} and \eqref{y22}  as a future work.

Next, we investigate the full time evolution of the complexity of formation for the holographic superconductor. We first have developed a general method to compute the {\it time-dependent} complexity of formation in section \ref{genmethod}.  By this method, we have computed the time dependent complexity of formation for the holographic superconductor, of which results are shown in Fig. \ref{TimeDepComResult11}.

 In both normal and superconducting  phase, the complexity of formation increases as time goes on. Furthermore, it increases linearly in time in the late time limit, which can be seen in Fig. \ref{TimeDerivativeResult1}.
In other words, the growth rate of the complexity is constant in the late time limit. The lower the temperature is, the smaller the growth rate of the complexity is. Furthermore, we compare the growth rate of the complexity in the late time limit with the Lloyd's bound, defined by the ADM mass.
If the temperature is high ($T \gg T_{c}$), the growth rate of the complexity saturates to the Lloyd's bound \cite{Lloyd_2000, Cai:2016xho, Yang:2016awy}. This is consistent with the result for the AdS-Schwarzschild Black brane case. As the temperature goes down, the growth rate of the complexity becomes smaller and smaller than the Lloyd's bound. Thus the complexity of the holographic superconductor does not violate the Lloyd's bound.

{While the Lloyd's bound was first introduced in the study of the CA conjecture, recent studies~\cite{Carmi:2017jqz,Kim:2017qrq,HosseiniMansoori:2018gdu,Ageev:2019fxn} show that the CA conjecture violates the Lloyd's bound in many cases. It was for the first time reported by \cite{Carmi:2017jqz,Kim:2017qrq} independently that even in the late time limit the CA conjecture violates the the Lloyd's bound because it saturates to the bound from above. However, based on our observation for the CV conjecture in this paper, we may propose a conjecture: {\it under a few `suitable' conditions on matter fields or causal structures of spacetime, the complexity by the CV conjecture may always satisfy the Lloyd's bound.}
It will be interesting to clarify what `suitable' means in the conjecture. We leave it as a future work.}

It will be also interesting to study the complexity of formation and the time-dependent complexity for other models including additional matter fields such as axion or dilaton fields.  Moreover, the complexity of the system at finite magnetic field or with higher derivative gravity model is other important directions to explore. In this paper, we focused on the CV conjecture only, but it will be interesting to study the complexity of superconductor by the holographic CA conjecture or other field theoretic methods, and compare them with our result. We leave these issues as future works.


\acknowledgments
We would like to thank Ioannis Papadimitriou for valuable discussions and correspondence.
The work of K.-Y. Kim and H.-S. Jeong was supported by Basic Science Research Program through the National Research Foundation of Korea(NRF) funded by the Ministry of Science, ICT $\&$ Future Planning(NRF- 2017R1A2B4004810) and GIST Research Institute(GRI) grant funded by the GIST in 2019. C. Niu is supported by the Natural Science Foundation of China under Grant No. 11805083.
 We also would like to thank the APCTP(Asia-Pacific Center for Theoretical Physics) focus program,``Holography and geometry of quantum entanglement'' in Seoul, Korea for the hospitality during our visit, where part of this work was done.


\appendix

\section{{Mass of the black hole in asymptotically AdS spacetimes}}\label{Defmass}
It is a subtle issue how to define the total mass of the black hole in asymptotically AdS spacetimes. In this appendix, we deal with three definitions of mass (the ADM mass, the Komar mass and the mass based on the holographically  renormalized stress tensor) and give their expressions in superconducting phase. In normal phase, all these three definitions are the same. but in superconducting phase, they are different. Thus, the ``mass'' in the Lloyd's bound is ambiguous.

{
All these three definitions need to deal with divergences. In particular, we may need to consider matter-dependent counterterms for complex scalar fields~\cite{Papadimitriou:2005ii,Papadimitriou:2007sj,Caldarelli:2016nni}. In our case, as the spacetime is static and the boundary is flat, the counterterm for a scalar field in all three cases have the following form
\begin{equation}\label{matterct1}
  S_{\text{ct}}=\int_{z=\epsilon}\sqrt{\gamma}\left[(c_1+c_2\ln\epsilon)|\phi|^2+\text{subleading temrs}\right]\td^3 x\,.
\end{equation}
Here  $c_1$ and $c_2$ are two constants, which depend on the mass (conformal dimension) and the charge of a scalar field. They are also different for three different masses. $\gamma$ is the determinant of the induced metric at the cut-off boundary $z=\epsilon\rightarrow0$.  With the metric~\eqref{anstz1} and the asymptotic expression~\eqref{asymfchi}, we obtain that
\begin{equation}\label{dergamma}
  \sqrt{\gamma}=e^{\gamma_0/2}\epsilon^{-3}(1+f_0\epsilon^3+\chi_1\epsilon^{3+\sqrt{4m^2+9}}+\cdots)\,.
\end{equation}
As we choose the boundary condition $\phi_+=0$, the scalar field has asymptotic expression $\phi\rightarrow\phi_-z^{(3+\sqrt{4m^2+9})/2}$. See Eq. \eqref{asymphi1}. Taking all these into account, we find that the leading term of the integrand  reads
\begin{equation}\label{leadingSct1}
  (c_1+c_2\ln\epsilon)|\phi_-|^2\epsilon^{\sqrt{4m^2+9}}\rightarrow0\,,
\end{equation}
where $m^2>-9/4$ (the BF bound).
Thus, there is no nonzero counterterms for matter fields in all these three masses.\footnote{{If we choose another boundary condition $\phi_-=0$ for a scalar field, the leading term of \eqref{matterct1} may be nonzero so the counterterms for a scalar field is needed. We also need to consider  finite boundary terms that impose
the relevant boundary conditions. For more details we refer to~\cite{Papadimitriou:2005ii,Papadimitriou:2007sj,Caldarelli:2016nni}.}}
}
\subsection{Generalized ADM mass}
The first one is the generalized ADM mass. This definition is the generalization of the ADM mass in asymptotically flat spacetimes proposed by Ref.~\cite{ABBOTT198276}.  In this paper, we deal with the black brane with a complex scalar hair. Ref.~\cite{ABBOTT198276} deals with the vacuum Einstein's equation without a cosmological constant but the generalization is straightforward. 

The generalized ADM mass is defined as conserved quantities constructed from the gravitational energy-momentum tensor and the Killing vectors $\bar{\xi}^\mu$ of a particular background metric tensor field $\bar{g}_{\mu\nu}$ which is the vacuum solution of Eq.~\eqref{eom3}.\footnote{In general, the vector $\bar{\xi}^\mu$ may not be the Killing vector of the real metric $g_{\mu\nu}$.} In the asymptotically AdS spacetimes, it is natural to choose $\bar{g}_{\mu\nu}$ as the pure AdS solution. Let us seperate the metric into two parts,
\begin{equation}\label{twometric}
  g_{\mu\nu}=\bar{g}_{\mu\nu}+h_{\mu\nu}\,,
\end{equation}
where we assume that $h_{\mu\nu}$ vanishes at infinity. Let us denote the covariant derivative corresponding to $\bar{g}_{\mu\nu}$ by $\bar{\nabla}_\mu$. To
facilitate this decomposition,  we will use the convention in this subsection that all operations such as index moving or differentiation are with respect to $\bar{g}_{\mu\nu}$. We can sperate the left hand side of Eq.~\eqref{eom3} into three pieces, i.e., the piece independent of $h_{\mu\nu}$ which vanishes because $\bar{g}_{\mu\nu}$ itself satisfies the vacuum Einstein's equation, the piece linear in $h_{\mu\nu}$ and the piece containing all quadratic and higher order terms in $h_{\mu\nu}$. Then we can write Eq.~\eqref{eom3} as
\begin{equation}\label{Einsteq2}
  R^{\mu\nu}_L-\frac12\bar{g}^{\mu\nu}R_L+\Lambda h^{\mu\nu}=\frac{1}{2}(\bar{T}^{\mu\nu}+ T^{\mu\nu})=T_{\text{total}}^{\mu\nu}\,.
\end{equation}
Here $-\bar{T}^{\mu\nu}/2$ contains all the quadratic and higher order terms of $h_{\mu\nu}$ in the left hand side of Eq.~\eqref{eom3}. The subscript $L$ refers to the terms linear in $h_{\mu\nu}$. Physically, $\bar{T}^{\mu\nu}$ is the energy-momentum tensor of the gravitational field and $T_{\text{total}}^{\mu\nu}$ is the total energy-momentum. Because the left hand side of Eq.~\eqref{Einsteq2} obeys the background Bianchi identity $R^{\mu\nu}_L-\frac12\bar{g}^{\mu\nu}R_L+\Lambda h^{\mu\nu}$ can lead that
\begin{equation}\label{dTeq1}
  \bar{\nabla}_\mu T_{\text{total}}^{\mu\nu}=0\,.
\end{equation}
Note that in Eq.~\eqref{dTeq1} the derivative is a background covariant derivative not an ordinary derivative. To construct a conserved current corresponding to Eq.~\eqref{dTeq1}, we can use the background Killing vector field $\bar{\xi}^\mu$ which satisfies that $\bar{\nabla}_\mu\bar{\xi}_\nu+\bar{\nabla}_\nu\bar{\xi}_\mu=0$. Then one can check that the combination $\xi_\nu T_{\text{total}}^{\mu\nu}$ is a conserved current and the corresponding conserved charge is
\begin{equation}\label{conversedE}
  E_A[\bar{\xi}]:=-\frac1{16\pi}\int_\Sigma\td\bar{\Sigma}_\nu\xi_\mu T^{\nu\mu}_{\text{total}}\,.
\end{equation}
Here $\Sigma$ is the $t=$constant hypersurface and $\td\bar{\Sigma}_\nu$ is induced volume element (corresponding to $\bar{g}_{\mu\nu}$). Now according to Sec.2 in Ref.~\cite{ABBOTT198276}, Eq.~\eqref{Einsteq2} can be expressed as
\begin{equation}\label{Tmunus1}
  T_{\text{total}}^{\mu\nu}=\bar{\nabla}_\alpha\bar{\nabla}_\beta K^{\mu\alpha\nu\beta}+X^{\mu\nu}\,.
\end{equation}
Here $K^{\mu\alpha\nu\beta}$ is called superpotential, which is defined as
\begin{equation}\label{superpot1}
  K^{\mu\alpha\nu\beta}:=\frac12(\bar{g}^{\mu\beta}H^{\nu\alpha}+\bar{g}^{\nu\alpha}H^{\mu\beta}-\bar{g}^{\mu\nu}H^{\alpha\beta}-\bar{g}^{\alpha\beta}H^{\mu\nu}) \,,
\end{equation}
with
\begin{equation}\label{defineHmunu}
  H^{\mu\nu}:=h^{\mu\nu}-\frac12\bar{g}^{\mu\nu}{h^\alpha}_\alpha\,.
\end{equation}
We see that the superpotential has the same index symmetry as the Riemann tensor. The additional term $X^{\mu\nu}$ can be written as,
\begin{equation}\label{auxilX1}
  X^{\mu\nu}:=-\frac12\bar{R}_{\alpha\beta\lambda}^{~~~~\nu}K^{\mu\lambda\alpha\beta}\,.
\end{equation}
Here $\bar{R}_{\alpha\beta\lambda}^{~~~~\nu}$ is the Riemann tensor defined by $[\bar{\nabla}_\alpha,\bar{\nabla}_\beta]v_\lambda=\bar{R}_{\alpha\beta\lambda}^{~~~~\nu}v_\nu$. Taking Eq.~\eqref{Tmunus1} into Eq.~\eqref{conversedE}, one can find that the right-hand side of Eq.~\eqref{conversedE} can be written as the surface integral. We then define the total ADM mass as
\begin{equation}\label{conservE2}
  E_A[\bar{\xi}]:=-\frac1{8\pi}\oint_S\td \bar{S}_{\mu\nu}(\bar{\nabla}_\alpha K^{\mu\nu\beta\alpha}-K^{\mu\alpha\beta\nu}\bar{\nabla}_\alpha)\bar{\xi}_\beta\,.
\end{equation}
Here $S$ is the codimension 2 surface at timelike infinity and $t=$constant and $\td \bar{S}_{\mu\nu}$ is the induced surface element (measured by $\bar{g}_{\mu\nu}$). Note Eq.~\eqref{conservE2} is not obtained from  Eq.~\eqref{conversedE} straightforwardly by the Guass's theorem as the inner boundary of the $\Sigma$ has been dropped. If we take that $\bar{\xi}^\mu=(\partial/\partial t)^\mu$, we obtain the ADM mass $M_{\text{ADM}}:=E_A[(\partial/\partial t)^\mu]$.

Now let us find the expression for the ADM mass under the ansatz~\eqref{anstz1}. The background metric $\bar{g}_{\mu\nu}$ and $h_{\mu\nu}$ can be expressed as
\begin{equation}\label{backmetric1}
  \bar{g}_{\mu\nu}=\frac1{z^2}\eta_{\mu\nu}\,, \qquad  h_{tt}=h_{zz}=f_0z^{d-2}+\cdots\,, \qquad h_{\mu\nu}=0~\text{for others}\,.
\end{equation}
After some algebras, one can finally find that,
\begin{equation}\label{ADMmass1}
  M_{\text{ADM}}=\frac{(d-1)\Sigma_{d-1}f_0}{16\pi}\,.
\end{equation}

\subsection{Generalized Komar mass}
For an asymptotically AdS spacetimes with a Killing vector $\xi^\mu$ we can define aother mass called the Komar mass, which can be treated as the generalization of the Komar integral in flat spacetime~\cite{Komar:1963aa}. The original Komar integral is divergent in the asymptotically AdS spacetime due to the cosmological constant. In order to cancel this divergence, Refs.~\cite{Barnich:2004uw,Kastor:2009wy} introduced a Killing potential $\omega^{\mu\nu}$ which is antisymmetric and satisfies the equation $\nabla_\nu\omega^{\mu\nu}=\xi^\mu$. The Killing potential exists locally because the Killing vector field $\xi^\mu$ satisfies $\nabla_\mu\xi^\mu=0$.

To construct the Komar mass, let us first define a current by $J^\mu:=\xi_\nu R^{\mu\nu}+(d-2)\xi^\mu/\el^2$. By the Bianchi identity, we have
\begin{equation}\label{dJR1}
  \nabla_\mu J^\mu=R^{\mu\nu}\nabla_\mu\xi_\nu+\xi_\nu\nabla_\mu R^{\mu\nu}+\frac{d-2}{\el^2}\nabla_\mu\xi^\mu=\frac12\xi^\mu\nabla_\mu R=0\,.
\end{equation}
The last equality is because the derivative of the Ricci scalar vanishes along the Killing vector field. We can now define the conserved quantity which is the energy associated with this current,
\begin{equation}\label{Komarxi1}
  E_K[\xi]:=-\frac1{4\pi}\frac{d-1}{d-2}\int_\Sigma\td\Sigma_\mu J^\mu\,.
\end{equation}
Using the Einstein's equation, one can check that this integration can be  expressed in terms of the energy momentum tensor field
\begin{equation}\label{Komarxi3}
  E_K[\xi]=-\frac{d-1}{d-2}\int_\Sigma\td\Sigma_\mu \left[T^{\mu\nu}-\frac{T}{d-1}g^{\mu\nu}\right]\xi_\nu\,.
\end{equation}

On the other hand, using the Killing equation $\nabla_\mu\xi_\nu+\nabla_\nu\xi_\mu=0$ and the fact that $\omega^{\mu\nu}$ is antisymmetric, we have
\begin{equation}\label{JRxi1}
  J_R^\mu=\xi_\nu R^{\mu\nu}=\nabla_\nu\left[\nabla^\mu\xi^\nu-\frac{d}{\el^2}\omega^{\mu\nu}\right]\,.
\end{equation}
This means the integrand in Eq.~\eqref{Komarxi1} is a total divergent term so the integral can be converted into a surface integration. We define the total Komar mass as
\begin{equation}\label{Komarxi2}
  E_K[\xi]:=-\frac1{16\pi}\frac{d-1}{d-2}\oint_S\td S_{\mu\nu} \left[\nabla^\mu\xi^\nu-\frac{d}{\el^2}\omega^{\mu\nu}\right]\,.
\end{equation}
Similarly, we have dropped the surface integration at the inner boundary of $\Sigma$ here. If we take that $\bar{\xi}^\mu=(\partial/\partial t)^\mu$, we obtain the ADM mass $M_{\text{Komar}}:=E_K[(\partial/\partial t)^\mu]$.

Now let us give the formula to compute the Komar mass under the metric ansatz~\eqref{anstz1}. The nonzero component of the Killing potential for the Killing vector field $(\partial/\partial t)^\mu$ is
\begin{equation}\label{Kpotential1}
  \omega^{tz}=H(z)e^{\chi(z)/2}z^{d+1}\,, \qquad H(z):=H_0-\int_{z_h}^ze^{-\chi/2}z^{-d-1}\td z\,.
\end{equation}
for an arbitrary constant $H_0$. Here $z_h$ is the position of horizon and $z_h=\infty$ for pure AdS spacetime. We set $H_0=0$ so the pure AdS spacetime has the zero Komar mass. Take Eq.~\eqref{Kpotential1} into the formula~\eqref{Komarxi2}, we obtain
\begin{equation}\label{KomarM}
  M_{\text{Komar}}=-\frac{\Sigma_{d-1}}{16\pi}\frac{d-1}{d-2}\lim_{z\rightarrow0^+}[z^{3-d}e^{\chi/2}(fe^{-\chi}z^{-2})'+2d H/\el^2]\,.
\end{equation}
Taking the asymptotic solutions of $f(z)$ and $\chi(z)$ in Eq.~\eqref{asymphi1} into account we finally have
\begin{equation}\label{KomarM2}
  M_{\text{Komar}}=\frac{(d-1)\Sigma_{d-1}f_0}{16\pi}+\frac{\Sigma_{d-1}}{16\pi}\frac{d-1}{d-2}\int^{z_h}_0e^{-\chi/2}z^{-d}\chi'\td z\,.
\end{equation}
We see that $M_{\text{Komar}}\neq M_{\text{ADM}}$ when $\chi(z)\neq0$ in general. Considering the first equation in Eq.~\eqref{eqsfchiAphi}, we find that $M_{\text{Komar}}\geq M_{\text{ADM}}$.

\subsection{Mass method based on holographic renormalized stress tensor}
In the framework of AdS/CFT correspondence, there is another important manner to define the mass. In this manner, the mass is defined by the boundary stress tensor at AdS boundary~\cite{Brown:1992br,Balasubramanian1999,Myers:1999aa}. Let us review how to obtain the mass by this method and then give the formula to compute it for the s-wave holographic superconductor.

Considering an asymptotically AdS spacetime with a cut off timelike boundary $\Sigma$, we can write the total action Eq.~\eqref{SetupModel} as
\begin{equation}\label{actionK1}
  I=I_{\text{bulk}}+\frac1{8\pi}\int_{\Sigma}\td^{d}x\sqrt{|h|}K\,,
\end{equation}
since there is no null boundary in this case. The quasi-local stress tensor $\mathcal{T}^{ab}$ on the surface $\Sigma$ is then defined through the variation of the action with respect to the boundary metric $h_{ab}$ and a suitable counterterm $S_{ct}$
\begin{equation}\label{stressT1}
  \mathcal{T}^{ab}:=\frac{2}{\sqrt{-h}}\left(\frac{\delta I}{\delta h_{ab}}+\frac{\delta S_{ct}}{\delta h_{ab}}\right)=\frac1{8\pi}(K^{ab}-Kh^{ab})+\frac{2}{\sqrt{-h}}\frac{\delta S_{ct}}{\delta h_{ab}}\,.
\end{equation}
The counterterm $S_{ct}$ is added because of divergences at the AdS boundary. Ref.~\cite{Emparan:1999aa} has shown how to construct the counterterms up to seven dimensional case. The expression for the boundary stress tensor has been computed up to five dimensional case in Ref.~\cite{Balasubramanian1999}.

For a background solving the equations of motion, this stress tensor will satisfy~\cite{Brown:1992br}
\begin{equation}\label{DstressT1}
  D_a\mathcal{T}^{ab}=-n_\mu T^{\mu\nu}{h_\nu}^b\,,
\end{equation}
where the source on the right-hand side is a projection of the matter stress-energy, $n_\mu$ is dual unit normal vector of $\Sigma$, and $D_a$ is the covariant derivative projected onto $\Sigma$. Under the source free condition for a complex scalar field, the right-hand side of Eq.~\eqref{DstressT1} will decay to zero at the AdS boundary, so we obtain a divergent free stress tensor at the boundary. Now if there is a boundary Killing vector $\xi^a$ such that $\mathcal{L}_\xi h_{ab}=0$, then we can construct a conserved current $\mathcal{T}^{ab}\xi_b$ at the boundary. Then we can define the mass contained at any time slice $S$ in $\Sigma$ as
\begin{equation}\label{BYKM}
  E_{h}[\xi]=\int_S\td S_a\mathcal{T}^{ab}\xi_b\,.
\end{equation}
Here $\td S_a$ is the induced volume element in $S$. If we take  $\xi^a=(\partial/\partial t)^a$,  we obtain the mass in this manner $M_{\text{h}}:=E_h[(\partial/\partial t)^a]$.

Now let us find the expression of $M_h$ for the metric ansatz~\eqref{anstz1}. According to Ref.~\cite{Balasubramanian1999}, the boundary stress tensor can be expressed as
\begin{equation}\label{stressT2}
  \mathcal{T}^{ab}=\frac1{8\pi}(K^{ab}-Kh^{ab}-\frac{d-1}{\el}h^{ab})\,.
\end{equation}
As the boundary is flat and only the first order counterterm is needed for all the cases with dimension $d>2$ we obtain
\begin{equation}\label{BYKM2}
  M_{\text{h}}=\frac{(d-1)\Sigma_{d-1}f_0}{16\pi}=M_{\text{ADM}}\,.
\end{equation}

\bibliographystyle{JHEP}

\providecommand{\href}[2]{#2}\begingroup\raggedright\endgroup

\end{document}